\documentclass[preprint]{aastex631}

\pdfoutput=1
\usepackage{bm,amsmath}
\shortauthors{Cheng et al.}

\graphicspath{{./}{figures/}}

\begin{document}

\title{Simulation of solar energetic particle events originated from coronal mass ejection shocks  with a data-driven physics-based transport model}

\author[0000-0002-3824-5172]{Lei Cheng}
\affiliation{Department of Aerospace, Physics and Space Sciences, Florida Institute of Technology, 150 W. University Blvd.
	Melbourne, FL 32901, USA}

\author[0000-0003-3529-8743]{Ming Zhang}
\affiliation{Department of Aerospace, Physics and Space Sciences, Florida Institute of Technology, 150 W. University Blvd. 
	Melbourne, FL 32901, USA}

\author[0000-0002-2106-9168]{Ryun Young Kwon}
\affiliation{Korea Astronomy and Space Science Institute, Daedeokdae-ro 776, Yuseong-gu Daejeon 34055, Republic of Korea}

\author[0000-0002-3176-8704]{David Lario}
\affiliation{ NASA, Goddard Space Flight Center, Heliophysics Science Division, 8800 Greenbelt Rd. Greenbelt, MD 20771, USA}

\correspondingauthor{Lei Cheng}
\email{lcheng@fit.edu}

\begin{abstract}

Solar energetic particle (SEP) events are associated with coronal mass ejections (CMEs)  and/or solar flares. SEPs travel through the corona and interplanetary space to reach Earth, posing a radiation hazard to spacecraft and astronauts working in space and the electronics on spacecraft. Due to the distinct magnetic field  configuration and solar eruption kinematic properties associated with each event, the utilization of a data-driven model becomes essential for predicting SEP hazards. In this study, we use a developed model that utilizes photospheric magnetic field measurements and  CME shock observations as inputs to simulate several historical SEP events associated with fast CME speeds ($>$700  $\rm km\;s^{-1}$). The model includes an SEP source term aligned with the theory of diffusive shock acceleration by the CME shock. The performance of the model is accessed by comparing simulations and observations of SEP intensity time profiles at SOHO, ACE, STEREO-A and STEREO-B.  The results generally matched observations well, particularly for protons below 40.0 MeV. However, discrepancies arose for higher-energy protons, notably for the events on 2011 March 7 and 2014 February 25, where the simulation tended to overestimate the proton flux .  At STEREO-A, the modeled proton intensities for the SEP events on 2013 April 11 and 2011 March 7  display a very different behavior compared to observations because of the efficient transport in longitude caused by the weak magnetic field.

\end{abstract}


\section{Introduction} \label{sec:intro}

Understanding the origin and transport of solar energetic particles (SEPs) is of paramount importance due to the risks posed by the exposure of spacecraft electronics and astronauts in space to high levels of energetic particles, including electrons, protons, and heavy ions up to GeV energy. Although the precise physical mechanism responsible for the acceleration of SEPs remains a subject of debate, it is widely accepted that SEPs are generated through two distinct processes: magnetic reconnection in solar flares and particle acceleration at shocks driven by coronal mass ejections (CMEs) \citep{baring1997diffusive}. In the past, SEP events have been categorized into impulsive SEP events, associated with solar flares, and gradual SEP events, which are believed to involve particle acceleration by CME-driven shocks. This paper focuses on simulating SEPs accelerated by shocks, as gradual SEP events typically exhibit higher proton intensities, longer durations, and  wider longitudinal spreads compared to those observed in impulsive events.

Over the decades, physics-based numerical models have been developed to investigate the propagation and acceleration of SEPs from their source to heliospheric spacecraft \citep[e.g.,][]{heras1992influence,heras1995three,Kallenrode1993, Bieberetal1994, Droge1994, NgReames1994, Ruffolo1995, KallenrodeWibberenz1997,lario1998energetic, zank2000particle, giacalone2000small, Ngetal2003, Riceetal2003, Lietal2003, Lee2005, qin2006effect, zhang2009propagation, Drogeetal2010, Luhmannetal2010, Kozarevetal2013, Marshetal2015, Huetal2017, zhang2017precipitation, Wijsen2022}. Although in-situ measurements and remote-sensing observations have significantly enhanced our understanding of the processes involved in the SEP event development, a model of the magnetic field and plasma in the region where SEPs are accelerated and transported is still necessary to simulate the production and propagation of SEPs.

In most previous models, the generation of SEPs was treated as an ad hoc input. For instance, energetic particles were often injected at a fixed radial distance in the corona, assuming an energy spectrum. However, SEPs produced by shocks driven by propagating CMEs are far more complex. Firstly, CME shocks can continuously accelerate particles at varying radial distances as they move away from the Sun. Secondly, the properties of CME shocks, such as their temporal evolution, radial distance, and front characteristics, exhibit considerable variability. Therefore, incorporating observed shocks as moving particle sources is crucial for SEP models, whenever feasible. Models for space weather prediction should realistically account for the propagation of CME shocks in the corona and interplanetary medium.

Our newly developed SEP model \citep{Cheng_2023,Zhang_2023} builds upon the SEP model proposed by \cite{zhang2009propagation} by incorporating the injection of source particles at the shock front location, which is reconstructed from coronagraph observations using an ellipsoid model \citep{kwon2014new}. This approach enables us to capture realistic conditions of the CME shock. To determine the intensity level of SEP particles, we employ a SEP seed injection model that utilizes input from shock properties. The application of our SEP model to a specific SEP event that occurred on 2020 May 29, involving a slow CME with a speed of approximately $\sim$300 $\rm km\;s^{-1}$, shows good agreement between the simulation results and observations \citep{Cheng_2023}. However, it should be noted that the weak CME shock driven by the slow CME in \cite{Cheng_2023} was limited to radial distances below 2.5 solar radii ($R_S$), and the SEPs were only accelerated up to approximately 16 MeV. In contrast, faster CMEs typically generate longer-lasting and stronger shocks, in principle capable of accelerating SEPs to higher energies. Therefore, to comprehensively evaluate the performance of our code, this paper includes simulations of several historical SEP events associated with faster CME speeds. Additionally, we improve the description of CME shock propagation beyond the last frame of observations by replacing the simple friction model utilized in \cite{Cheng_2023} with the analytic model described by \citet{2013SoPh..285..391C,2017SpWea..15..464C}.

The structure of this paper is as follows: In Section \ref{sec:model}, we provide a detailed description of our simulation model, which employs time-backward stochastic differential equations to derive the time-intensity profile of SEPs at specific locations in the heliosphere. Section \ref{sec:res} presents the simulation results for several historical SEP events. Finally, Section \ref{sec:disc} offers a discussion and summary of our results.

\section{Model Description} \label{sec:model}

 We adopt our newly developed solar energetic particle (SEP) model \citep{Cheng_2023,Zhang_2023},a time backward stochastic three-dimensional (3D) focused transport simulation of SEPs that expanded the SEP model introduced by \cite{zhang2009propagation}. It is built on integrating a source particle injection at the CME shock front position. The shock location is determined using an ellipsoid model reconstructed from coronagraph observations \citep{kwon2014new}.
 
The SEP model incorporates the coronal magnetic field spanning from 1 $\rm R_S$ to 2.5 $\rm R_S$ by employing the potential field source surface (PFSS) model based on synoptic magnetogram observations taken just after the eruption \citep{zhang2017precipitation}. The use of time-dependent synoptic magnetograms does not significantly alter the results \citep{Zhao_2018} because the overall distribution of SEPs is primarily influenced by large-scale magnetic fields.  We adopt a reference frame co-rotating with the Sun with stationary coronal magnetic field. The SEP distribution functions are set to zero at the absorbed inner boundary of 1 $\rm R_s$ and at the outer boundary of 20 au.
The heliospheric magnetic field beyond 2.5 $\rm R_S$ is obtained from a Parker spiral model with empirical solar wind speed and density profiles \citep{leblanc1998tracing}. 

\begin{figure}
	\epsscale{1.1}
	\plotone{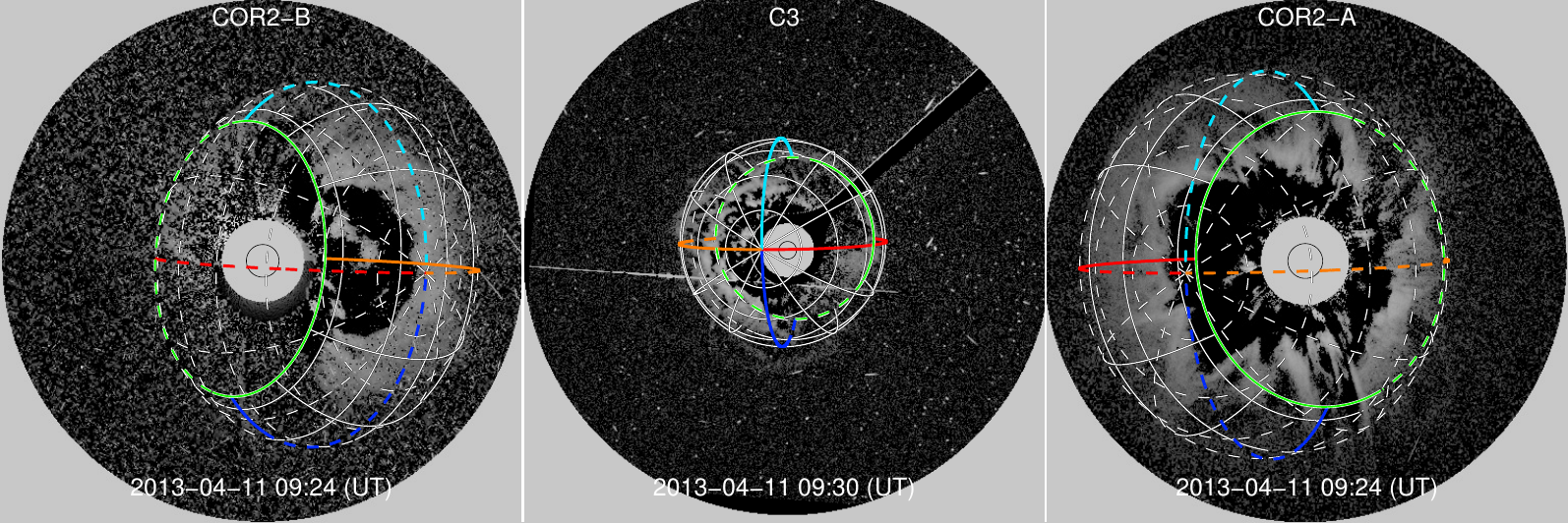}
	\caption{Coronagraph observations of the 2013 April 11 CME from three vantage locations at \textit{STEREO-B}, \textit{SOHO}, and \textit{STEREO-A} (from left to right) and their fits to a 3D
		ellipsoid model. Overplotted is the ellipsoid model that represents the 3D geometry of the shock. Red, orange, blue, and cyan lines refer to the four quadrants of the
		ellipsoid. White lines are used to indicate the surface of the ellipsoid.
		\label{fig:2013apr11event}}
\end{figure}

The sources of accelerated SEPs comove with the CME shock. The location, shape, and temporal evolution of the CME are determined using an ellipsoid model developed by \cite{kwon2014new}. A three-dimensional CME shock surface is reconstructed by using EUV and white-light coronagraph images obtained from instruments on \textit{STEREO}, \textit{SDO}, and \textit{SOHO}. These observations cover a radial range from a few $\rm R_S$  to tens of $\rm R_S$.  Figure \ref{fig:2013apr11event} shows a sample of the reconstructed CME shock structure overplotted on coronagraph images from left to right, \textit{STEREO-B}, \textit{SOHO}, and \textit{STEREO-A}. 

This construction can be done up to the time when the CME shock goes beyond field the view (FOV) of the coronagraphs on board the spacecraft. However, when the leading point of the shock is already out of the FOV, a model is required to extend the propagation of the CME shock.	
In this study, we have adopted the analytic model described by \cite{2013SoPh..285..391C,2017SpWea..15..464C} for CME propagation. Their research reveals that initially, the fast CME propagates at approximately a constant speed ($V_{0 \mathrm{cme}}$) and drives a shock. Once the CME reaches a certain distance, it starts to decelerate and becomes decoupled from the shock, while the shock maintains a constant speed. Eventually, both the CME and its associated shock decelerate. Thus, the speed of the shock front beyond the last available frame can be expressed as given by \cite{2013SoPh..285..391C}.
	
	\begin{equation}\label{shockmodel}
	V_{\mathrm{shf}}(t)= \begin{cases}V_{0 \mathrm{cme}}, & t<\tau_{c } \\ \left(V_{0 \mathrm{cme}}-V_{1 \mathrm{AU}}\right)\left(\frac{t}{\tau_{c }}\right)^{-1 / 3}+V_{1 \mathrm{AU}}, & t \geq \tau_{c }\end{cases},
	\end{equation}
	where $V_{1 \mathrm{AU}}$ represents  the solar wind speed at 1 au, and $\tau_{c }$ denotes the critical time when the shock begins to decelerate, and $V_{0 \mathrm{cme}}$ is the initial speed of the CME. The angular size of CMEs available on the DONKI website \footnote{\url{https://kauai.ccmc.gsfc.nasa.gov/DONKI/search/}\label{donki}}   is utilized to constrain the size of the shock. Linear interpolation is employed to estimate the angular size between the last observed frame of the CME shock and the time it reaches 21.5 $R_S$. 
	
	  To determine the properties of the shock, we input relevant parameters (plasma adibatic index $\gamma$, upstream magnetic obliquity $\theta_{b n 1}$, Alfv\'en Mach number  $M_{A 1}$, and fast magnetosonic Mach number $M_{MS 1}$) into the MHD shock adiabatic equation\citep[e.g.,][]{Thompson1962, Kabin2001JPlPh..66..259K} to calculate the plasma compression ratio $R$ at the shock. The downstream plasma properties (density, velocity, magnetic field, and plasma thermal speed) are determined using the general jump conditions for oblique MHD shocks \citep{fitzpatrick2014plasma}, focusing solely on solutions for fast-mode shocks.
	
	 The particles are injected at the shock from the thermal tail of solar wind ions with the Maxwellian velocity distribution of downstream solar wind ions since the thermal tail particles downstream of the shock with sufficient energy behave more likely as seed particles for diffusive shock acceleration. A particle injection speed $v_{inj}= a_0 V_{shf}/cos(\theta_{b n 1})$ is adopted with $a_0$ being a constant \citep{Lee_2005,le_Roux_2009}. A  range of $a_0$ between $2.3-2.7$ generally produces a good match with observations through our investigations. 
		
	 Determining the cutoff momentum $p_c$  of the shock power-law distribution is essential for calculating the proton intensity. We apply diffusive shock acceleration time $t_{acc}$  below \citep{Drury1983} to obtain the cutoff momentum via setting $t_{acc}$ as the minimum between the shock lifetime $t$ and the adiabatic particle energy loss time  $t_{loss}$ in the background solar wind plasma with $t_{loss} = 3 (\nabla \cdot {\bf V})^{-1}$.
	\begin{equation}\label{acctime}
		t_{acc}=\int_{p_{inj}}^{p_c} \frac{3}{V_1-V_2} \left [ \frac{\kappa_1}{V_1}+\frac{\kappa_2}{V_2} \right] \frac{dp}{p},
	\end{equation}
	where $\kappa_1$ and $\kappa_2$ denotes the upstream and downstream the particle diffusion coefficients, respectively, with $p_{inj}$ being the momentum of injected seed particles.

	 In the simulation, we assume a constant radial mean free path, $\lambda_r$, for particles with a rigidity of 1 GV. 
	We assume perpendicular diffusion below is mainly driven by field line random walk \citep{zhang2017precipitation}. 
	\begin{equation}
	\kappa_\bot = \frac{v}{2V} k \kappa_{gd0} \frac{B_0}{B} ,
	\end{equation}
	where the terms $v/V$ and  $B_0/B$ denotes the ratio of particle speed to solar wind plasma speed and the expansion of the magnetic field flux tube from the solar surface, respectively, with $\kappa_{gd0} = 3.4 \times 10^{13}$ cm$^2$ s$^{-1}$ being the diffusion coefficient in the photosphere.  We introduce the factor $k$ to moderate the transmission of field line diffusion from the photosphere to the corona.  In this simulation,  we set  $k=0.074$ and $\lambda_r$= 50 $R_S$ for the event on 2013 April 11. We set $k=0.074$  and  $\lambda_r$= 200 $R_S$ for the events on 2011 March 7 and 2011 November 3. For the 2014 February 25 event, we set $k=0.0074$ and $\lambda_r$= 10 $R_S$.

\section{Results} \label{sec:res}
We proceed to apply our model calculation to simulate several SEP events associated with fast CMEs, as documented in the SOHO/LASCO CME catalog\footnote{\url{https://cdaw.gsfc.nasa.gov/CME_list/index.html}\label{lasco}}  \citep{2004JGRA..109.7105Y, 2009EM&P..104..295G}. These specific SEP events have been selected due to the availability of CME imagery from three view points (\textit{STEREO-A} (STA), \textit{SOHO} and \textit{STEREO-B} (STB)) that allowed us to reconstruct with enough fidelity the large-scale structure of CME shocks.

\subsection{2013 April 11 SEP Event}
\begin{figure}
	\epsscale{0.9}
	\plotone{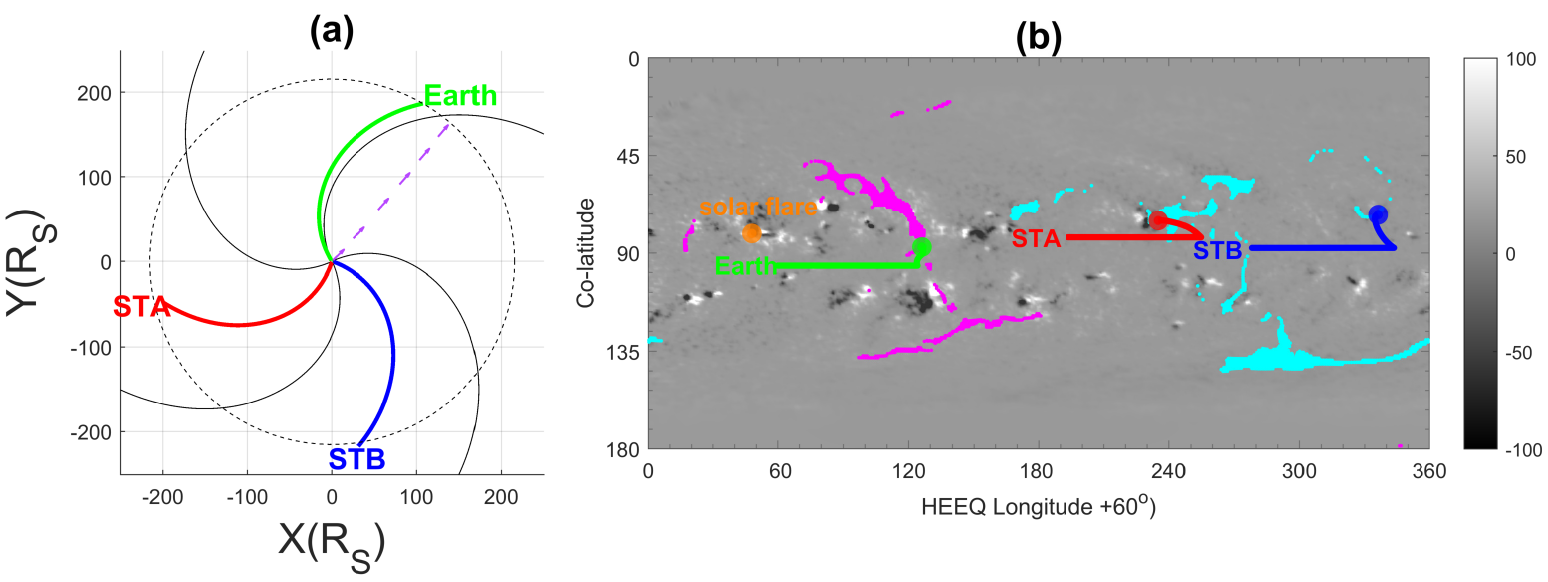}
	\caption{(a) Locations of \textit{STEREO-A}(STA), \textit{Earth} and \textit{STEREO-B} (STB) on 2013 April 11 in the equatorial plane as seen from the north, where $R_S$ is the solar radius. The purple arrow denotes the moving direction of the CME shock; The blue, green, and red curves are the Parker spiral and coronal magnetic field lines that connect to the spacecraft.   The panel (b) shows the projection on the solar surface of \textit{STEREO-A}(STA), \textit{Earth} and \textit{STEREO-B} (STB). The gray background image is the strength of the measured photospheric magnetic field. The cyan and magenta dots are the inward and outward open field lines. The red, green and blue dots mark the footpoint of the field lines connecting to \textit{STEREO-A}, \textit{Earth} and \textit{STEREO-B}, respectively. The orange dot is the site of solar flare. A shift ($60^{\circ}$) of the heliocentric Earth equatorial (HEEQ) coordinate system in longitude is used. Earth is at $60^{\circ}$ in longitude.
		\label{fig:2013apr11location}}
\end{figure}

The 2013 April 11 SEP event has been investigated in a study by \citet{2014ApJ...797....8L}. Here, we provide a brief summary of the observational findings. On 2013 April 11 (day of year 101), \textit{STEREO-A} was at a heliocentric radial distance of 0.96 AU and 134$^{\circ}$ west of Earth in terms of longitude, while \textit{STEREO-B} was located approximately 141$^{\circ}$ behind Earth and at a heliocentric radial distance of 1.02 AU, as depicted in Figure \ref{fig:2013apr11location}(a). The associated solar flare occurred at N07E13 in the heliocentric Earth equatorial (HEEQ) coordinate system. Prior to this event, the period was characterized by relative calmness. The solar flare started around 06:55 UT on 2013 April 11, reaching its peak intensity at 07:16 UT, estimated to be at the level of an M6.5 class flare according to the DONKI website\textsuperscript{\ref{donki}}. Additionally, a halo CME with a speed of 861 $\rm km\;s^{-1}$ was reported in the LASCO CME Catalog\textsuperscript{\ref{lasco}}. The purple arrow in Figure \ref{fig:2013apr11location}(a) represents the direction of propagation of apex of the CME shock, reconstructed using an ellipsoid model \citep{kwon2014new} based on coronagraph observations.

Figure \ref{fig:2013apr11location}(b) shows a magnetogram obtained at 09:04UT on 2013 April 11 where the regions marked in the cyan and magenta colors indicate open field lines. We show the projection on the solar surface of \textit{STEREO-A} (STA), \textit{Earth} and \textit{STEREO-B} (STB). All spacecraft are connected to open field line regions. The footpoints of the field lines connecting to \textit{STEREO-A} (red dot), \textit{Earth} (green dot) and \textit{STEREO-B} (blue dot) are traced using Parker spiral up to a distance of 2.5 $R_S$ and below that coronal field lines from the PFSS model. The longitudinal distance between the footpoint of the field line connecting to \textit{Earth} and the location of the solar flare (orange dot) is approximately 78$^{\circ}$. Similarly, the longitudinal distance between the red dot and the orange dot is largest $\sim 187^{\circ}$, while the longitudinal distance between the blue dot and the orange dot is the smallest $\sim 72^{\circ}$. The solar wind speed  is 339 $ \rm km\;s^{-1}$, 529 $ \rm km\;s^{-1}$ and 391 $ \rm km\;s^{-1}$ at \textit{Earth}, \textit{STEREO-A} and \textit{STEREO-B}, respectively.

\begin{figure}
	\epsscale{0.9}
	\plotone{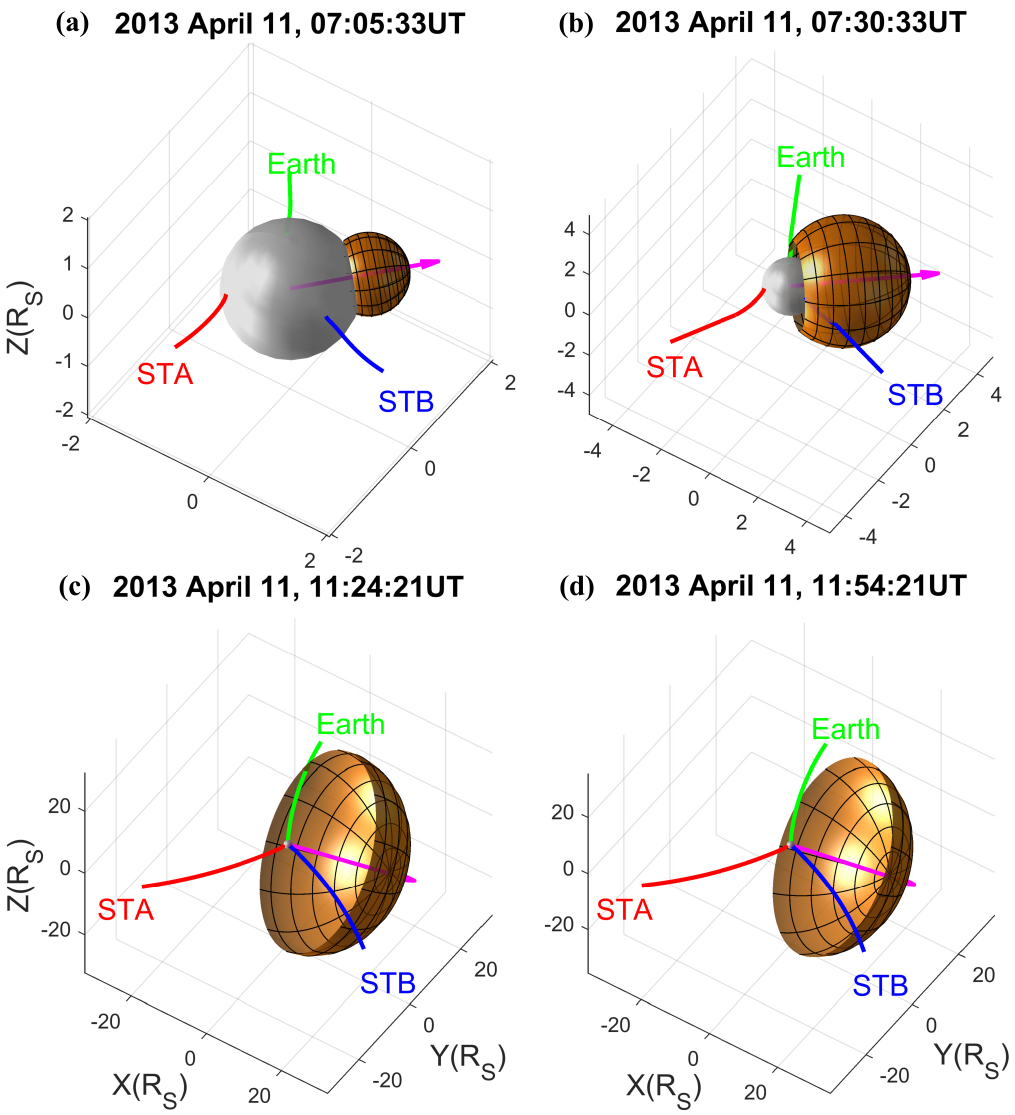}
	\caption{Time evolution of ellipsoid CME shock surfaces (orange color) on 2013  April 11  with magnetic field lines connecting to \textit{Earth} (green), \textit{STEREO-A} (STA) (red), and \textit{STEREO-B} (STB) (blue), where the Sun (gray sphere) is at the center. The purple arrow denotes the moving direction of the CME shock.
		\label{fig:2013apr11CME}}
\end{figure}

	Figure \ref{fig:2013apr11CME} illustrates the temporal evolution of ellipsoidal CME shock surfaces (depicted in orange) along with magnetic field lines connecting to \textit{Earth} (green), \textit{STEREO-A} (red), and \textit{STEREO-B} (blue). At the initial time of 07:05:33 UT (Panel a), none of the spacecraft were connected to the CME shock. However, around 07:30:33 UT(Panel b), both \textit{Earth} (green) and \textit{STEREO-B} established a connection with the CME shock. Subsequently, approximately 4 hours later(Panel c), \textit{STEREO-B} lost its connection to the shock, followed by \textit{Earth} losing its connection half an hour later(Panel d). Throughout the entire period, \textit{STEREO-A} did not establish magnetic connection with the reconstructed shock via field lines.

\begin{figure}
	\epsscale{1.0}
	\plotone{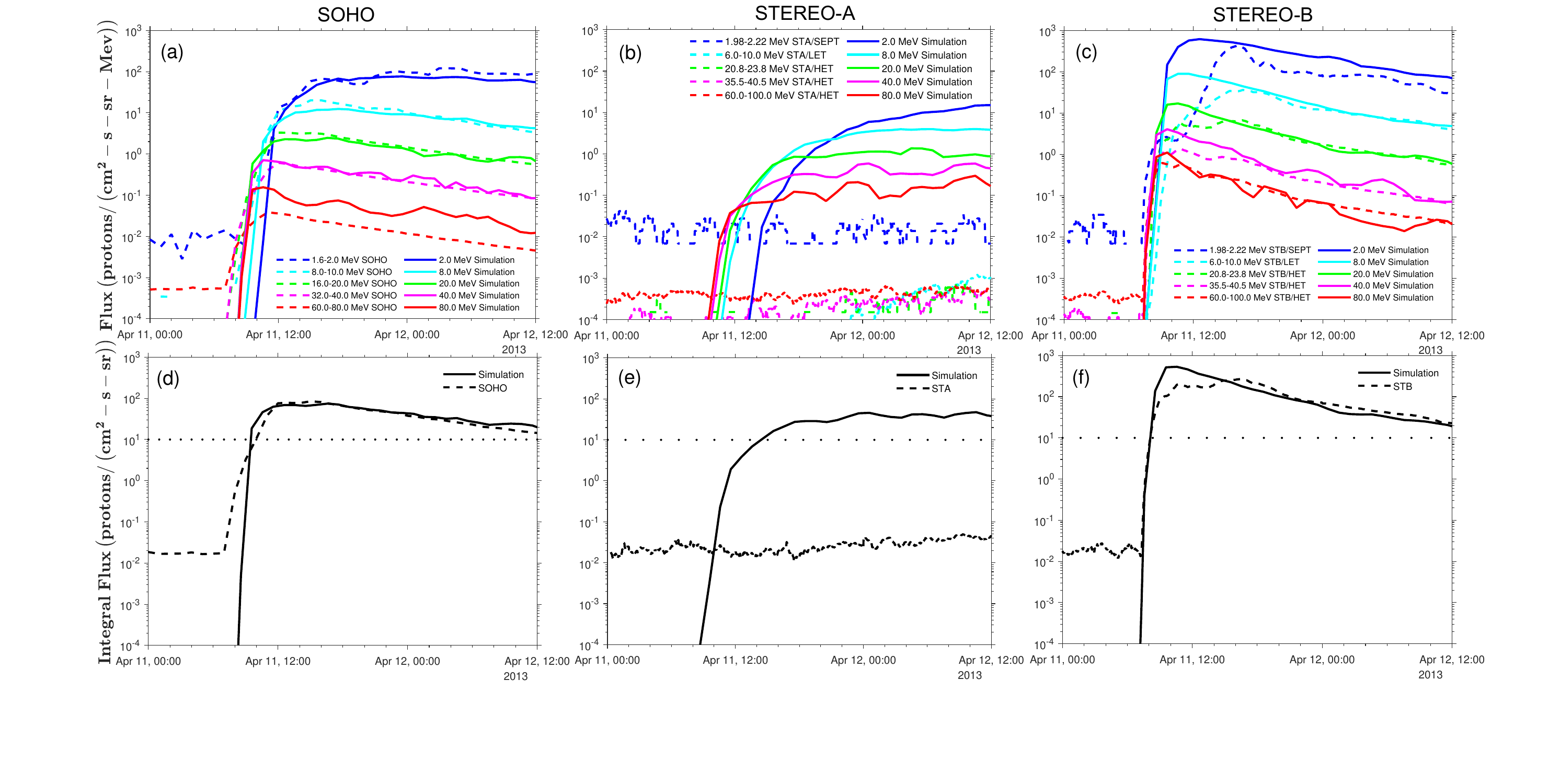}
	\caption{Time evolution of (Top) differential proton  flux on April 11, 2013 between 2.0 MeV and 80.0 MeV at the loci of (a) Earth, (b) \textit{STEREO-A} and (c) \textit{STEREO-B},  and (Bottom)  integral flux for $>$10.0 MeV protons at the location of (d) Earth, (e) \textit{STEREO-A} and (f) \textit{STEREO-B} from the simulation (solid curves) and  observations (dash curves). The dotted line denotes the threshold flux  (10 proton flux units ((cm$^{2}$ s sr)$^{-1}$) for energies $>$ 10 MeV)  of a SEP event in the traditional NOAA definition. Data of observations at Earth are from ERNE instrument of \textit{SOHO} near the Sun-Earth L1 point.  Data of observations at \textit{STEREO-A} and  \textit{STEREO-B}  are from the SEPT, LET and HET instruments of \textit{STEREO-A} and \textit{STEREO-B}. 
		\label{fig:2013apr11flux}}
\end{figure}

Figure \ref{fig:2013apr11flux} presents the observed proton intensity-time profiles at various energies (indicated by dashed curves) and corresponding integral flux for proton energies $>$ 10 MeV,   as measured by \textit{SOHO}, \textit{STEREO-A} and \textit{STEREO-B}, alongside the corresponding profiles obtained from our simulations (depicted by solid curves). It is worth noting that the \textit{SOHO} spacecraft is positioned in a halo orbit around the Sun-Earth L1 point, which is situated approximately 0.01 au away from Earth. Data used at Earth were obtained from SOHO/ERNE \citep{1995SoPh..162..505T}. 
Overall, the simulated proton flux exhibits a strong correlation with the observed data, capturing the evolution of the flux accurately.
The modeled differential proton flux (solid curves) at the location of Earth, ranging from 2.0 MeV to 40.0 MeV, closely matches the observed results (dash curves) measured by SOHO/ERNE within a factor of 2. However, for 80.0 MeV protons near the cut-off frequency, the modeled flux (red solid curve) peaks at $0.16$ (cm$^{2}$ s sr MeV)$^{-1}$, which is four times larger than the peak value of $0.038$ (cm$^{2}$ s sr MeV)$^{-1}$ observed for 60.0-80.0 MeV protons (red dash curve). This indicates that our simulation overestimates the proton intensities for protons with energies close to the cut-off frequency at Earth during this event.  The onset time of simulation results at Earth is around one and a half hours later than observations as shown in Figure \ref{fig:2013apr11flux}(d)

The rise in the proton flux observed by STEREO-B occurs at nearly the same time (right column in Figure \ref{fig:2013apr11flux}(c)). The observations from STEREO-B are obtained by the Solar Electron and Proton Telescope (SEPT) \citep{2008SSRv..136..363M} for the energy range of 1.98 MeV to 2.22 MeV, the Low-Energy Telescope (LET) \citep{mewaldt2008low} for 6.0 MeV to 10.0 MeV protons, and the High-Energy Telescope (HET) \citep{2008SSRv..136..391V} for protons between 20.8 MeV and 100.0 MeV. Unlike the flux of protons at energies higher than $\sim$ 30MeV, the flux at lower energies display two peaks. For example, the observed proton flux at 1.98 MeV to 2.22 MeV clearly shows two peak values: 2.6 (cm$^{2}$ s sr MeV)$^{-1}$ at 09:31 UT and 431.8 (cm$^{2}$ s sr MeV)$^{-1}$ at 16:14 UT. The onset time of the first peak occurs even earlier than that of the higher-energy SEPs. Considering that the proton strengths of impulsive SEP events are typically lower than those measured in gradual events, the first and second enhancements in the proton flux may be attributed to the acceleration of magnetic connections in solar flares and CME shocks, respectively. The modeled flux from our simulation agrees with the observed values in all energy channels during the decreasing phase after 16:14 UT. However, our simulation exhibits a relatively sharp increase in proton flux with  peak values up to three times larger than STEREO-B observations, probably due to an overestimation of the shock cutoff at its particle source locations.  The onset time of the simulation results at the location of \textit{STEREO-B} is almost the same as that from observations as shown in Figure \ref{fig:2013apr11flux}(f).

\begin{figure}
	\epsscale{1.4}
	\plotone{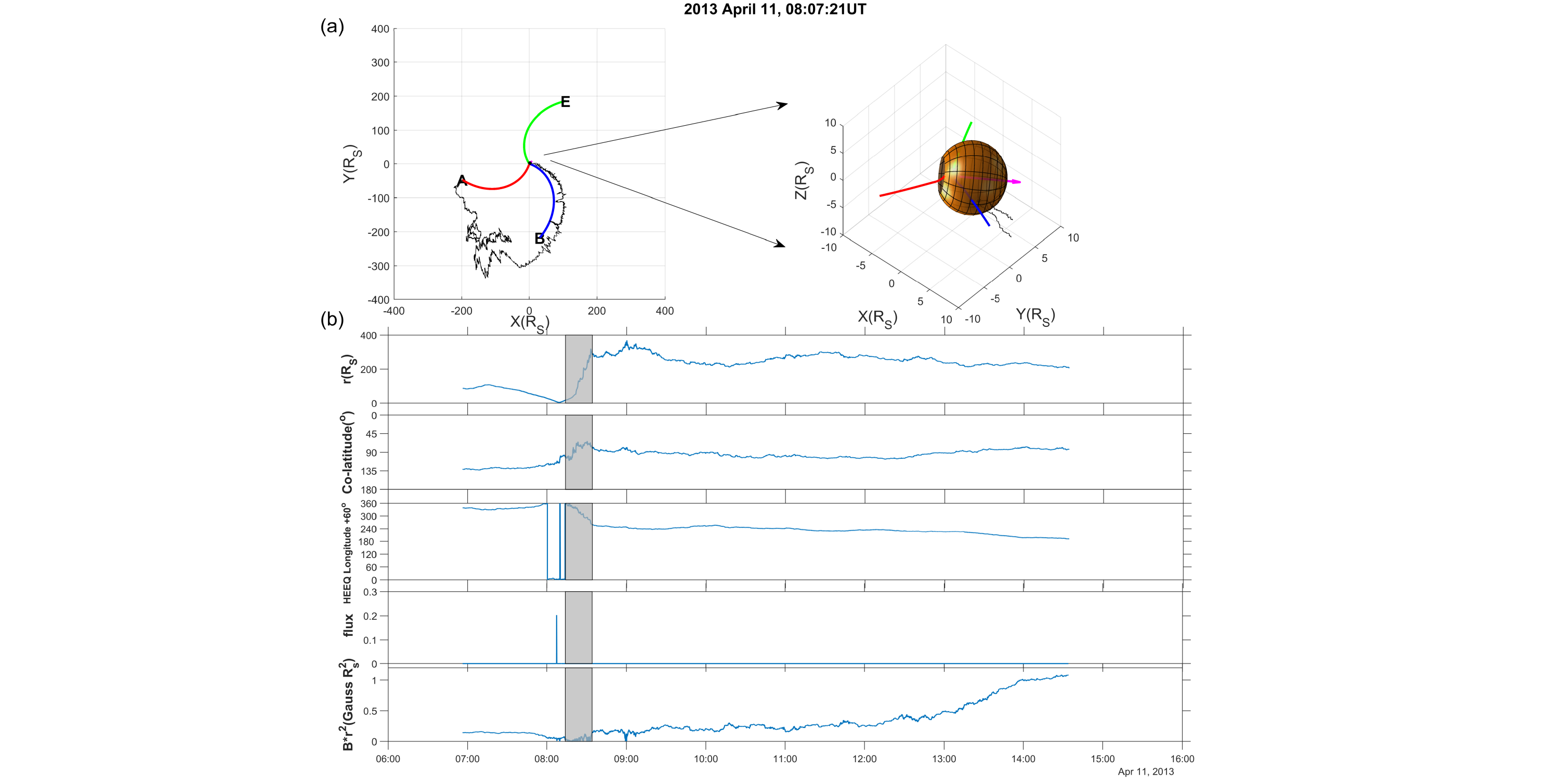}
	\caption{(a) Two different views of a typical path (black curves) of 8.0-MeV protons at \textit{STEREO-A} traced backward from  14:33 UT on 2013 April 11 with CME shock surfaces (orange color) at 08:07 UT on 2013 April 11  when the path crossed the shock surface. Locations of \textit{STEREO-A}(A), \textit{Earth}(E) and \textit{STEREO-B} (B) on 2013 April 11  in the equatorial plane are seen from the north. The purple arrow denotes the moving direction of the CME shock; the blue, green, and red curves are the Parker spiral and coronal magnetic field lines that connect to each spacecraft. (b) Time evolution of radial distance(r), Co-latitude, longitude (HEEQ longitude plus 60 degrees), intensity and normalized magnetic field (magnetic field strength multiplied by $r^2$) for the traced path.
		\label{fig:2013apr11path}}
\end{figure}

Unlike the consistent enhancement of modeled proton flux with observations at the loci of Earth and \textit{STEREO-B}, our calculations for STEREO-A show the enhancements of SEP proton fluxes above the background level of the instrument during the observed time period (middle column in Figure \ref{fig:2013apr11flux}). This result is not consistent with the absence of significant SEP enhancements in the observations from STEREO-A. To explain this inconsistency, a typical traced path of proton at 8.0 MeV traced backward from  14:33 UT on 2013 April 11 is shown in Figure \ref{fig:2013apr11path}. The proton undergoes diffusion and finally transport similar to the transport of proton at \textit{STEREO-B} as seen in the left plot of Figure \ref{fig:2013apr11path}(a). The left plot of Figure \ref{fig:2013apr11path}(a) shows the projection over the  equatorial plane of the path followed by a proton observed by STEREO-A. The intensity at \textit{STEREO-A} should be quite small since the magnetic field lines connected to \textit{STEREO-A} are rooted into a region far away from CME shock (oranged color) in the right plot of Figure \ref{fig:2013apr11path}(a). However the particle path does cross the CME shock twice at 08:07 UT and 08:11 UT on April 11 as seen by the zoom-in 3-D view of the right plot. That means large diffusion exists during the transport of protons. The large diffusion is attributed to the very weak local magnetic field strength from the PFSS model.
Figure \ref{fig:2013apr11path}(b) shows the time history of radial distance (r), Co-latitude, longitude (HEEQ longitude plus 60 degrees), intensity and normalized magnetic field (magnetic field strength multiplied by $r^2$) for the path traced by the particle. From 08:14UT to 08:34UT (gray shade regions), the accelerated source proton transported from around $348^o$  to $258^o$  in the longitude due to the weak magnetic field in the vicinity of the current sheet and finally arrived \textit{STEREO-A} around $191^o$  at  14:33 UT. This discrepancy between simulation and observations may be attributed to differences in the configuration of the actual magnetic field, which differs from the PFSS model used in the simulation. Simulation using different magnetic field models  with more realistic current sheets might resolve this issue.

Figure {\ref{fig:2013apr11_skf}  presents the time evolution of the wavefront properties at the leading point including the radial distance ($r_{shf}$), wavefront speed relative to the upstream plasma ($u_1$),the Alfv\'en speed ($V_A$), solar wind speed ($V_{sw}$), Alfv\'en Mach number ($M_A=u_1/V_A$), fast magnetosonic Mach number ($M_{MS}$), the shock compression ratio ($R$) and oblique angle ($\theta_{bn}$).  During the time period, the maximum value of the compression ratio is $\sim$3.2-3.8. Initially, the leading point of the shock front showed quasi -perpendicular conditions ($\theta_{bn} > 45^o$) then became quasi-parallel (Figure  {\ref{fig:2013apr11_skf}(e)). This results from the transition from the PFSS field configuration below $2.5 R_S$ to the radial field beyond that point.  Considering that the particle acceleration is calculated at the whole shock surface,  acceleration from quasi-perpendicular regions also exists at other positions of the shock surface as the shock moves away from the Sun. Note that Figure  {\ref{fig:2013apr11_skf} is only for the leading point (or apex) of the reconstructed shock. It is possible that other regions of the shock front have  different compression ratios and oblique angles.

\begin{figure}
	\epsscale{0.95}
	\plotone{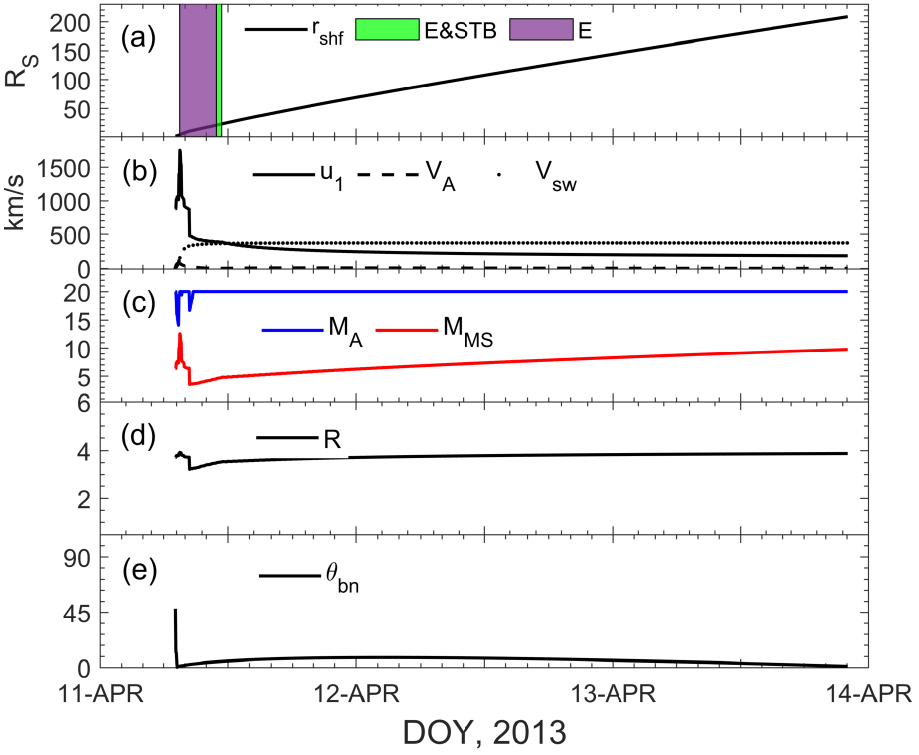}
	\caption{Time variation of (a) the radial distance; (b) the speed of wavefront ($u_1$) in the frame of solar wind, solar wind speed ($V_{sw}$) and Alfv\'en speed ($V_A$); (c) Alfv\'en Mach number ($M_A$), magnetosonic Mach number ($M_{MS}$), (d) compression ratio (R) and (e) oblique angle ($\theta_{bn}$) measured at the apex of the shock for 2013  April 11 SEP event. The color shade regions in the top panel indicate the time when Earth and STEREO-B are connected to the shock.
		\label{fig:2013apr11_skf}}
\end{figure}


\subsection{2011 March 7 SEP Event}
Now we applied our simulation to the 2011 March 7 SEP event \citep{Park_2013}. On March 7, 2011 (day of year 66), STEREO-A was located 88$^{\circ}$ west of Earth in longitude and at a heliocentric radial distance of 0.96 AU, while STEREO-B was approximately 95$^{\circ}$ behind Earth and at a heliocentric radial distance of 1.0 AU, as shown in Figure \ref{fig:2011mar7location}(a). The solar wind speed at these three locations ranged from approximately 300 to 450 $\rm km \;s^{-1}$. All of the spacecraft, including STEREO-A, SOHO, and STEREO-B, observed the enhancement of particle flux.

However, we should indicate that different injections occurred during this time interval. The main SEP injection occurred at around 20:00UT in association with  a M3.7 flare at N30W48 as seen from Earth. This solar flare began around 19:43 UT  and peaked at 20:12 UT.  A halo CME was observed starting at 20:00UT from the LASCO CME Catalog, with a speed of 2125 $\rm km\;s^{-1}$. The magenta arrows in Figure \ref{fig:2011mar7location}(a) represents the direction of motion for the reconstructed CME shock based on multipoint coronagraph observations using the ellipsoid model. Before this solar flare, there was a weaker flare located at N11E21 which began around 13:44 UT and peaked at 14:30 UT with an estimated intensity at the M2.0 class. The associated CME was observed starting at 14:48UT as reported in the LASCO CME Catalog, with a speed of 698 $\rm km\;s^{-1}$. This prior CME contributed with an earlier small particle intensity enhancement at \textit{STEREO-B} just before the main SEP intensity increase simulated here

Figure \ref{fig:2011mar7location}(b) depicts, with the same format as Figure \ref{fig:2013apr11location}(b), the projection on the solar surface of the positions of \textit{STEREO-A} (STA), \textit{Earth}, and \textit{STEREO-B} (STB). The footpoints of the magnetic field lines connecting to \textit{STEREO-A} (red dot), \textit{Earth} (green dot), and \textit{STEREO-B} (blue dot) are traced using Parker spiral and coronal field lines obtained from the PFSS model. The longitudinal distances between the footpoints of the field lines connecting to \textit{Earth} (green dot), \textit{STEREO-A} (red dot), \textit{STEREO-B} (blue dot), and the location of the solar flare (orange dot) are approximately $\sim 33^{\circ}$, $\sim 56^{\circ}$, and $\sim 77^{\circ}$, respectively.


\begin{figure}
	\epsscale{0.9}
	\plotone{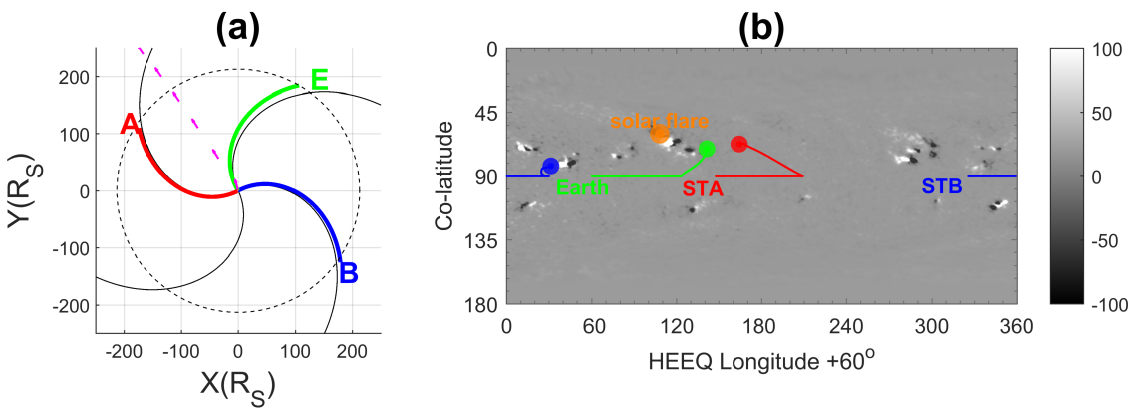}
	\caption{(a) Locations of \textit{STEREO-A}(STA), \textit{Earth} and \textit{STEREO-B} (STB) on 2011 March 7 in the equatorial plane as seen from the north, where $R_S$ is the solar radius. The magenta arrow denotes the moving direction of the CME shock; The blue, green, and red curves are the Parker spiral and coronal magnetic field lines that connect to the spacecraft.   The panel (b) shows the projection on the solar surface of \textit{STEREO-A}(STA), \textit{Earth} and \textit{STEREO-B} (STB). The gray background image is the strength of the measured photospheric magnetic field.  The red, green and blue dots mark the footpoints of the field lines connecting to \textit{STEREO-A}, \textit{Earth} and \textit{STEREO-B}, respectively. The orange dot is the site of solar flare. A shift ($60^{\circ}$) of the heliocentric Earth equatorial (HEEQ) coordinate system in longitude is used. Earth is at $60^{\circ}$ in longitude.
		\label{fig:2011mar7location}}
\end{figure}

Figure \ref{fig:2011mar7CME} illustrates the time evolution of the ellipsoid CME shock surface (orange structures) with magnetic field lines connecting to \textit{Earth} (green), \textit{STEREO-A} (red), and \textit{STEREO-B} (blue). Throughout the propagation of the CME shock, \textit{Earth} maintained a magnetic connection to the shock surface for the longest duration. Initially at 19:50:50 UT, none of the spacecraft were magnetically connected to the CME shock . Around 19:55:50 UT, \textit{Earth} established a magnetic connection with the shock surface in its western plank. \textit{STEREO-A} and \textit{STEREO-B} subsequently established magnetic connections with the reconstructed CME shock at 20:08:35 UT at the base of its eastern and western plank, respectively. One and a half hours later, at 21:38:35 UT, only \textit{Earth} kept a magnetic connection with the reconstructed CME shock. Finally, at 22:38:35 UT, none of the spacecraft were magnetically connected to the fitted CME shock.

\begin{figure}
	\epsscale{0.9}
	\plotone{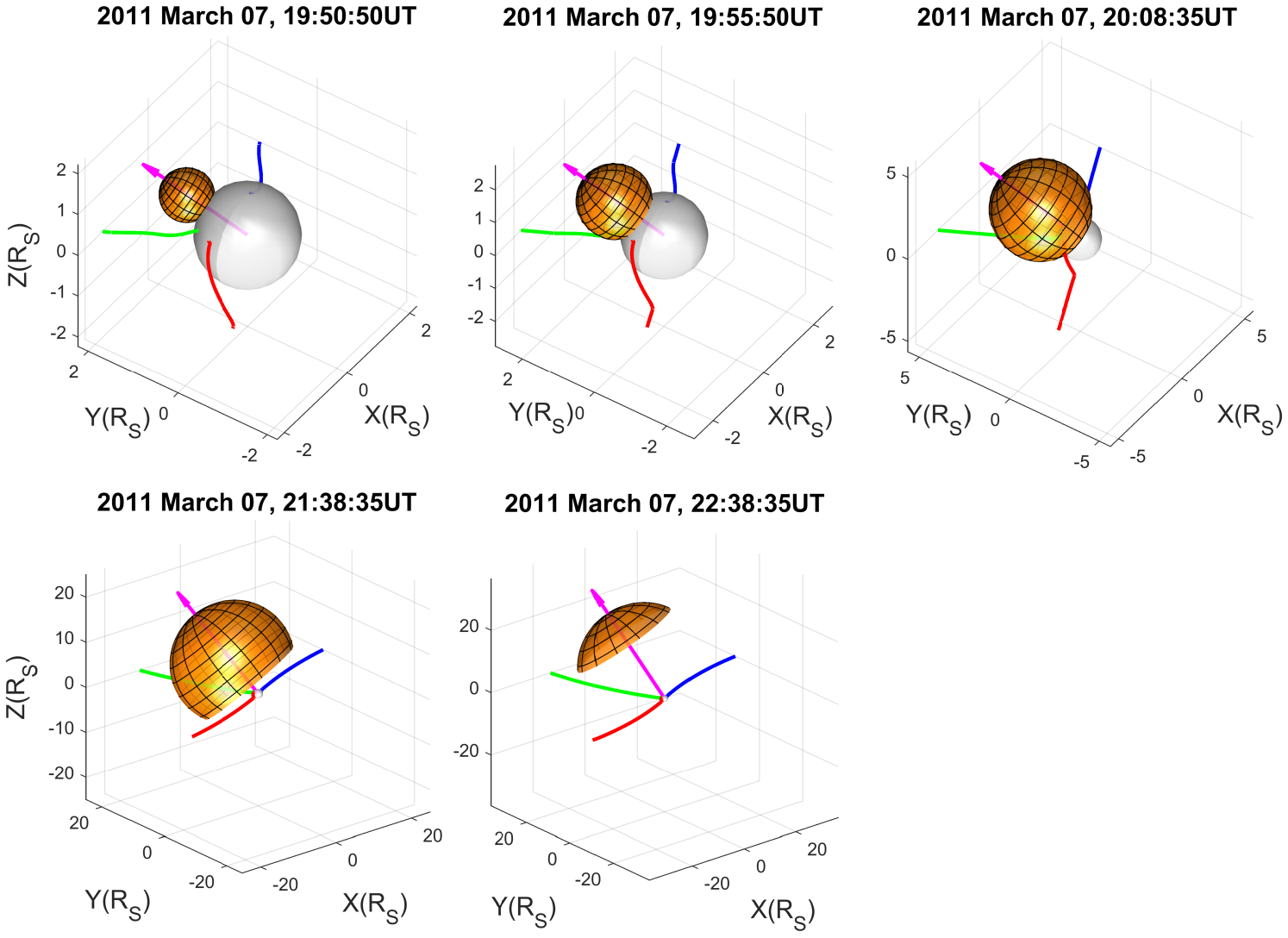}
	\caption{Time evolution of ellipsoid CME shock surfaces (orange color) on 2011 March 7  with magnetic field lines connecting to \textit{Earth} (green), \textit{STEREO-A} (STA) (red), and \textit{STEREO-B} (STB) (blue), where the Sun is at the center. The purple arrow denotes the moving direction of the CME shock.
		\label{fig:2011mar7CME}}
\end{figure}

Figure \ref{fig:2011mar7flux} displays the proton intensity-time profiles at different energies observed (open symbols) by \textit{SOHO}, \textit{STEREO-A}, and \textit{STEREO-B}, along with the results obtained from our simulations (solid traces).
There are two enhancements of flux during this SEP event with the first enhancement starting at around 15:00UT with a relatively smaller peak and the second one starting at around 20:00UT with a relatively much higher peak (see Figure \ref{fig:2011mar7flux}(a) and Figure \ref{fig:2011mar7flux}(c)). The first enhancement of proton flux is attributed to the acceleration of CME shock driven by the weaker prior CME starting at 14:48UT.
The second stronger CME is the one reconstructed here (Figure \ref{fig:2011mar7CME}) and the resulting simulated SEP intensities should be compared with the second observed particle intensity enhancement. Overall, the evolution of the proton flux from the simulation shows a good correlation with the second and main particle intensity increase observed by \textit{SOHO} and \textit{STEREO-B} especially for low-energy particles below 40 MeV. The simulated flux of protons at 2.0 MeV, 8.0 MeV, and 20.0 MeV exhibits variations that are generally within a factor of 2 compared to the observed values. However, for 40.0 MeV and 80.0 MeV protons, the modeled flux is approximately one and two orders of magnitude higher than the observed values, respectively. Similarly, the simulated flux of protons at 2.0 MeV and 8.0 MeV closely matches the observed values at \textit{STEREO-B} while the modeled flux of 80.0 MeV protons is around two orders of magnitude greater than the observed values. These results also suggest that our simulation tends to overestimate the flux of high-energy protons near the cut-off frequency due to an overestimation of shock cutoff energy. Thus, the modeled integral flux are higher than those observed by \textit{SOHO} and \textit{STEREO-B} as seen in Figure \ref{fig:2011mar7flux}(d) and Figure \ref{fig:2011mar7flux}(f). The estimated onset time is around one hour later than those actually observed.

\begin{figure}
	\epsscale{1}
	\plotone{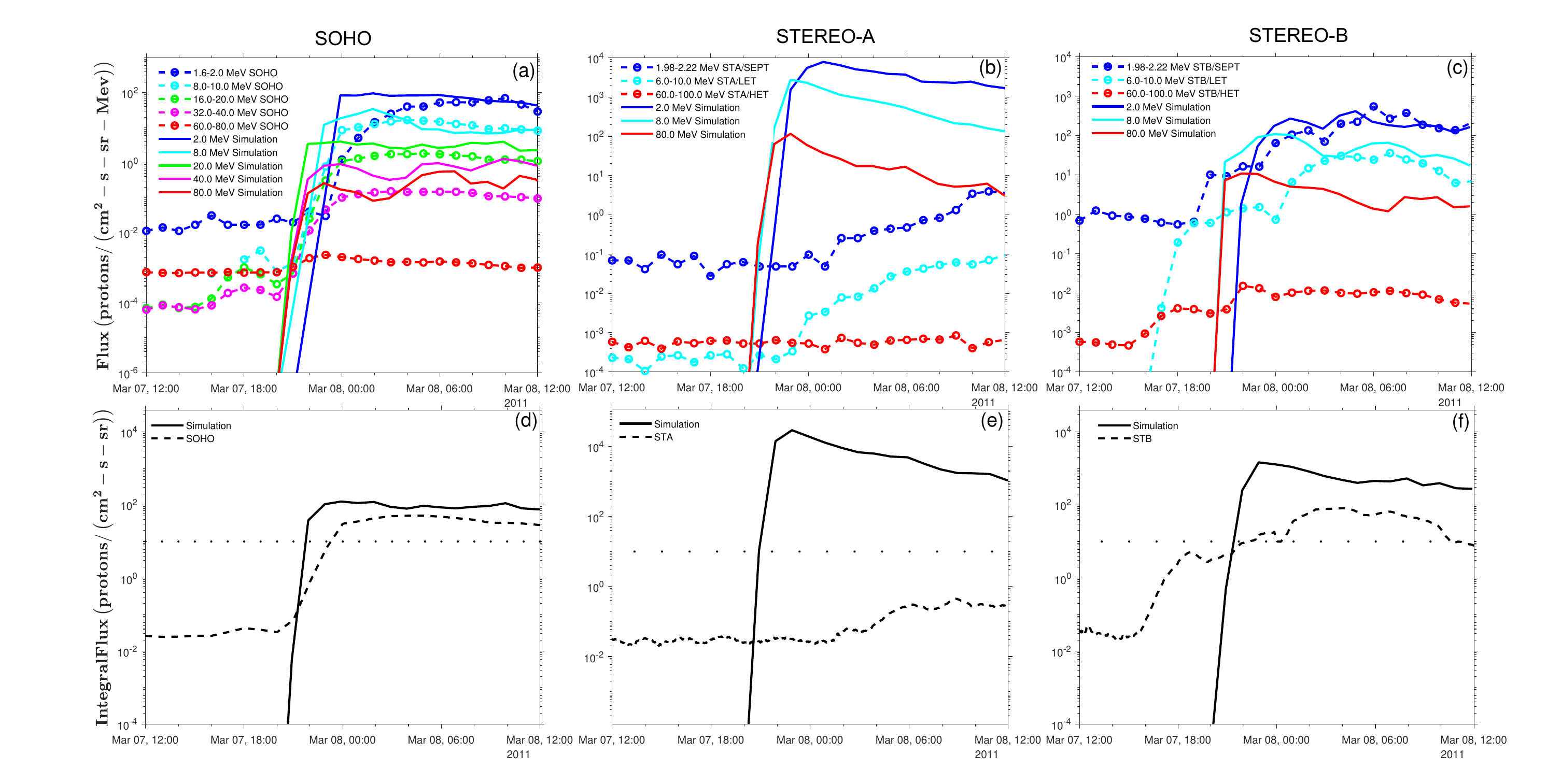}
	\caption{Time evolution of (Top) differential proton  flux on 2011 March 7 between 2.0 MeV and 80.0 MeV at the loci of (a) Earth, (b) \textit{STEREO-A} and (c) \textit{STEREO-B},  and (Bottom)  integral flux for $>$10.0 MeV protons at the location of (d) Earth, (e) \textit{STEREO-A} and (f) \textit{STEREO-B} from the simulation (solid curves) and  observations (dash curves). The dotted line denotes the threshold flux  (10 proton flux units ((cm$^{2}$ s sr)$^{-1}$) for energies $>$ 10 MeV)  of a SEP event in the traditional NOAA definition. Data of observations at Earth are from ERNE instrument of \textit{SOHO} near the Sun-Earth L1 point.  Data of observations at \textit{STEREO-A} and  \textit{STEREO-B}  are from the SEPT, LET and HET instruments of \textit{STEREO-A} and \textit{STEREO-B}. 
		\label{fig:2011mar7flux}}
\end{figure}

In contrast to the proton flux observed by \textit{SOHO} and \textit{STEREO-B}, the flux observed at \textit{STEREO-A} in Figure \ref{fig:2011mar7flux}(e)  exhibits a more gradual increase. Our modeled flux, on the other hand, shows a sharp rising phase that is not consistent with the observed gradual increase. Figure \ref{fig:2011mar7blines}(a) show the  contour plots of normalized magnetic field in the equatorial plane from 1 $R_S$ to 2.5  $R_S$  as seen from the north on 2011 March 7 with magnetic field lines connecting to \textit{Earth} (green), \textit{STEREO-A} (STA) (red), and \textit{STEREO-B} (STB) (blue). The dots mark the footpoints of \textit{Earth} (green dot), \textit{STEREO-A} (STA) (red dot), and \textit{STEREO-B} (STB) (blue dot) on $2.5 R_S$ traced using the Parker spiral with solar wind speed (left plot) $V_{sw}= 370$ $\rm km \;s^{-1}$ and (right plot) $V_{sw}= 320$ $\rm km \;s^{-1}$ . Black lines are magnetic field lines. Even though  the magnetic field lines change the polarity traced from STEREO-A when lowering the solar wind speed by $\sim$ 50 $\rm km\;s^{-1}$, there is little influence on the proton intensity as seen in the ime evolution of  differential proton  flux for 2.0 MeV protons at the location \textit{STEREO-A} in Figure \ref{fig:2011mar7blines}(b) . STEREO-A is well connected to the CME shock at the start of the event in our simulation, but the observations tell a different story. This discrepancy is due to the same reason as discussed in the inconsistent behavior of proton flux at STEREO-A for the 2013 April 11 SEP event, i.e., the large transport of protons  because of the very weak local magnetic field strength,  which might be resolved by adopting different magnetic field models. 

\begin{figure}
	\epsscale{1}
	\plotone{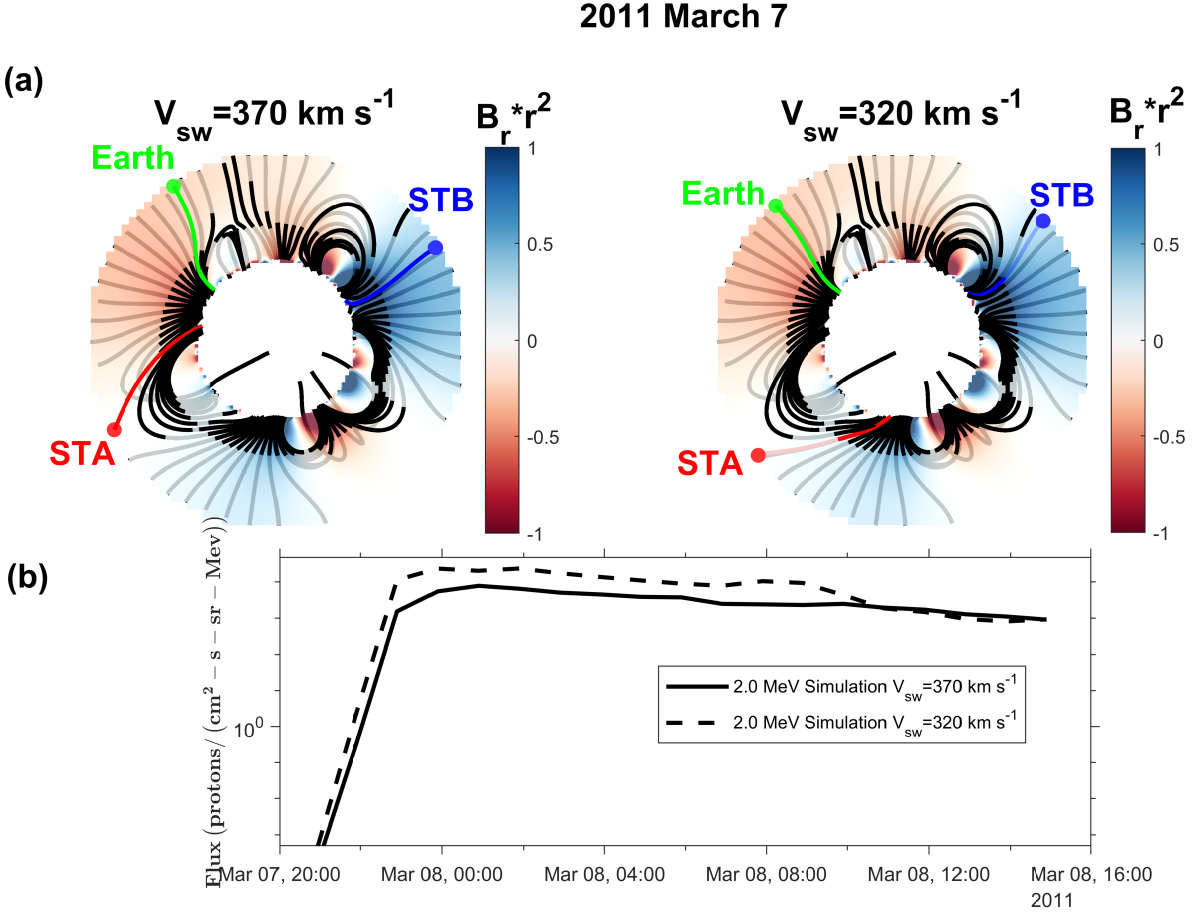}
	\caption{(a) Contour plots of normalized magnetic field (magnetic field strength multiplied by $r^2$) in the equatorial plane from 1 $R_S$ to 2.5  $R_S$  as seen from the north on 2011 March 7 with magnetic field lines connecting to \textit{Earth} (green), \textit{STEREO-A} (STA) (red), and \textit{STEREO-B} (STB) (blue). The dots mark the footpoints of \textit{Earth} (green dot), \textit{STEREO-A} (STA) (red dot), and \textit{STEREO-B} (STB) (blue dot) on $2.5 R_S$ traced using the Parker spiral with solar wind speed (left plot) $V_{sw}= 370$ $\rm km \;s^{-1}$ and (right plot) $V_{sw}= 320$ $\rm km \;s^{-1}$ . Black lines are magnetic field lines.  (b) Time evolution of  differential proton  flux on 2011 March 7 for 2.0 MeV protons at the location \textit{STEREO-A} for $V_{sw}= 370$ $\rm km \;s^{-1}$ (solid line) and $V_{sw}= 320$ $\rm km \;s^{-1}$ (dash line). 
		\label{fig:2011mar7blines}}
\end{figure}

\subsection{2011 November 3 Event}
The SEP event observed by \textit{STEREO-A}, \textit{STEREO-B} and near-earth spacecraft on 2011 November 3  has been investigated by several papers \citep[e.g.,][]{Park_2013,2014SoPh..289.1731P,Gomez_2015}.
On 2011 November 3 (day of year 307), there is a longitudinal separation of approximately 105° between STEREO-A and Earth, as well as a separation of approximately 102° between STEREO-B and Earth as shown in Figure \ref{fig:2011nov3_location}(a).  The solar conditions preceding this SEP event were characterized by a certain level of complexity. Multiple active regions (ARs) were observed including AR 11339 (N18E60) near the footpoint of STEREO-B, AR 11333 (N10W85) close to the Earth footpoint and  AR(N05E160)  on the backside of the Sun as seen from Earth\citep{Park_2013}. A halo CME with a speed of $\sim$ 1100 $\rm km\;s^{-1}$ was associated with an EUV emission observed near the backside AR \citep{2013SoPh..288..241N}. The EUV emission  was estimated to be located at N08E156 starting at approximately 22:11 UT and reaching its peak intensity at 22:41 UT, that would have been associated with a solar flare with an estimated magnitude falling within the range of M4.7 and X1.4 class.

This SEP event has been modeled by \citet{Zhang_2023} using an input of corona and heliospheric plasma and magnetic field configuration from a MHD model. Here we mainly try to find the difference of results between both simulations using PFSS model and MHD model as the input magnetic field configuration.

The time evolution of the ellipsoid CME shock surfaces (orange surfaces) with magnetic field lines connecting to \textit{Earth} (green), \textit{STEREO-A} (red), and \textit{STEREO-B} (blue) is illustrated in Figure \ref{fig:2011nov3_CME}. The CME shock surfaces are exactly the same CME surfaces used in \citet{Zhang_2023}. The duration of field lines connection to the CME shock surfaces are slightly different due to the different magnetic field configurations used in both work.  Initially,  only \textit{STEREO-A} was magnetically connected to the CME shock as shown in the plot at 22:24:16 UT on November 3. Unlike the brief connection $\sim$ 1 hour duration to the shock surfaces in \citet{Zhang_2023} for \textit{STEREO-B} at the earlier time before 01:24:16 UT on  November 4, no connection is established in our simulation. From 04:54:16 UT on  November 4, \textit{STEREO-A} no longer connected to the CME shock. Around two days and 10 hours later at 14:54:16 UT on November 6, \textit{STEREO-B} began to establish  magnetic connection with the shock surface already beyond the spacecraft radial distance , and this connection was maintained for 6.5 hours.     Finally, after 21:24:35 UT on November 6, none of the spacecraft were magnetically connected to the CME shock. Throughout the propagation of the CME shock, \textit{Earth} never established  magnetic connection to the shock surface.

Figure \ref{fig:2011nov3_flux} presents the proton intensity-time profiles at different energies observed (dash curves) by \textit{SOHO}, \textit{STEREO-B}, and \textit{STEREO-A}, along with the results obtained from our simulations (solid curves).
In spite SOHO never established magnetic connection with the reproduced shock surface, the simulation results exhibit SEP intensity enhancement at Earth. This result is due to cross-field diffusion transport of the SEPs.
The simulated flux of protons at energies 2.0 MeV, 8.0 MeV, and 15.0 MeV at exhibits variations that are generally within a factor of 2 of  the observed intensities. 
However, for protons above 40.0 MeV at \textit{SOHO} and \textit{STEREO-B}, the modeled flux is much lower than the observed values, which does not agree with the good match between observation and simulation in  \citet{Zhang_2023}. 
At \textit{STEREO-A}, the proton flux for all energy channels are lower than the observed ones, which agree with the result in \citet{Zhang_2023}. The lower modeled flux may be attributed to  the influence of shock-generated turbulence, which alters the diffusion coefficient along the magnetic field lines connected to STEREO-A \citep{Zhang_2023,Zhang:2023aL} or strong anisotropies at the onset observed by all the spacecraft \citep{Gomez_2015}.

Although the magnitude of modeled integral fluxes are lower than the observations, our simulation results do agree with the observations on forecasting whether a SEP event was observed.    The onset time of modeled integral flux at STEREO-A agrees with that from observations while the modeled onset time at \textit{SOHO} and \textit{STEREO-B} are later than the observations as seen in Figure \ref{fig:2011nov3_flux}(d), Figure \ref{fig:2011nov3_flux}(e) and Figure \ref{fig:2011nov3_flux}(f).

\begin{figure}
	\epsscale{0.9}
	\plotone{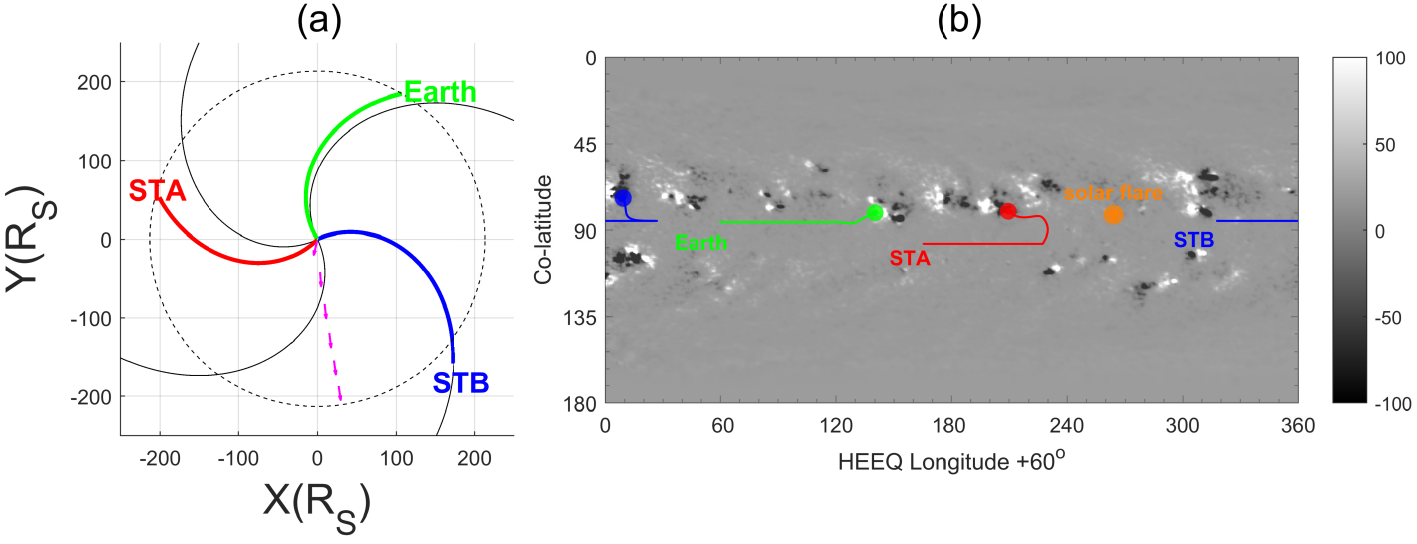}
	\caption{(a) Locations of \textit{STEREO-A}(STA), \textit{Earth} and \textit{STEREO-B} (STB) on 2011 November 3 in the equatorial plane as seen from the north, where $R_S$ is the solar radius. The magenta arrow denotes the moving direction of the CME shock; The blue, green, and red curves are the Parker spiral and coronal magnetic field lines that connect to the spacecraft.   The panel (b) shows the projection on the solar surface of \textit{STEREO-A}(STA), \textit{Earth} and \textit{STEREO-B} (STB). The gray background image is the strength of the measured photospheric magnetic field.  The red, green and blue dots mark the footpoints of the field lines connecting to \textit{STEREO-A}, \textit{Earth} and \textit{STEREO-B}, respectively. The orange dot is the site of solar flare. A shift ($60^{\circ}$) of the heliocentric Earth equatorial (HEEQ) coordinate system in longitude is used. Earth is at $60^{\circ}$ in longitude.
		\label{fig:2011nov3_location}}
\end{figure}

\begin{figure}
	\epsscale{0.9}
	\plotone{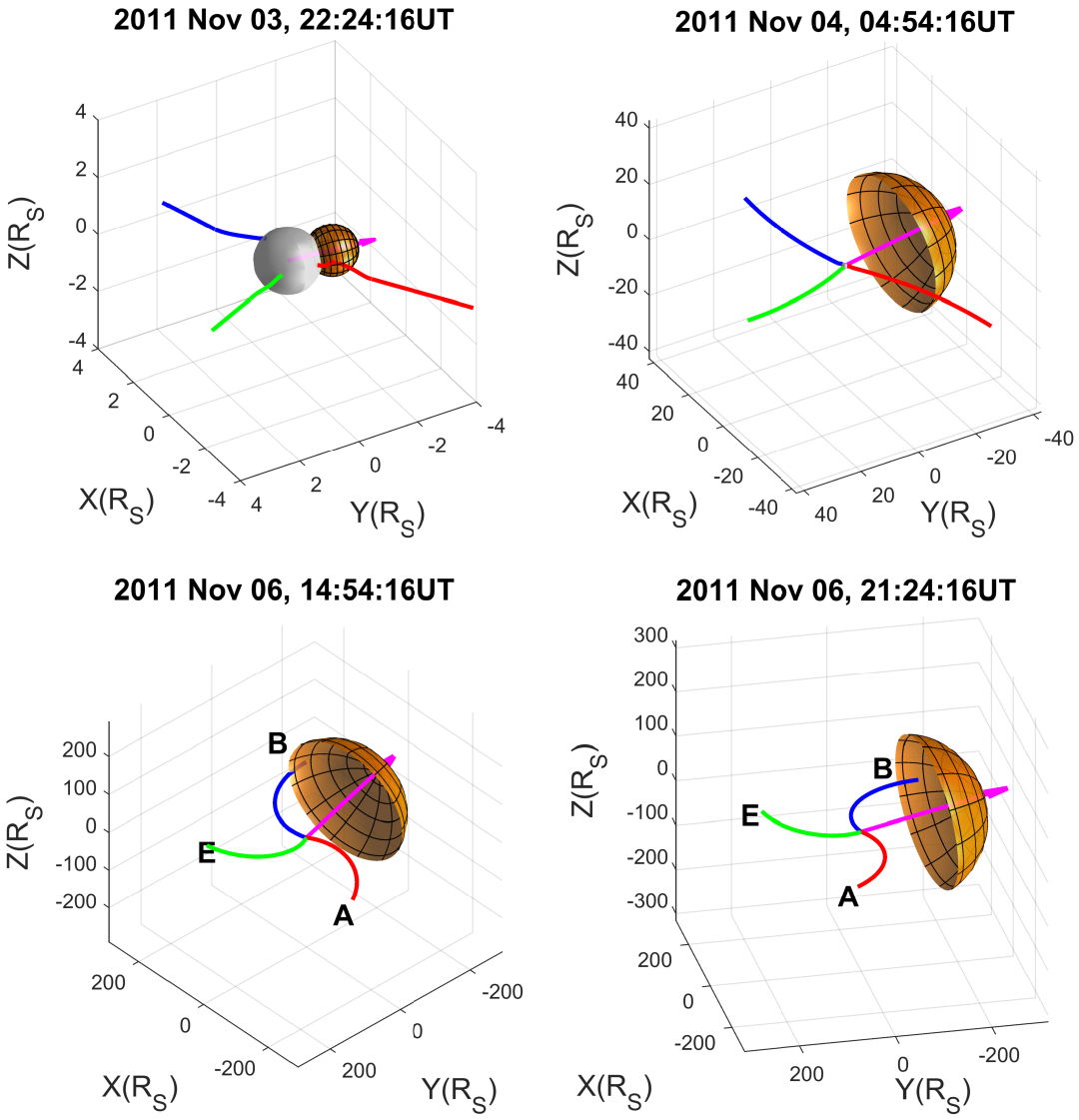}
	\caption{Time evolution of ellipsoid CME shock surfaces (orange color) for 2011 November 3 SEP event  with magnetic field lines connecting to \textit{Earth} (green), \textit{STEREO-A} (STA) (red), and \textit{STEREO-B} (STB) (blue), where the Sun is at the center. The purple arrow denotes the moving direction of the CME shock.
		\label{fig:2011nov3_CME}}
\end{figure}

\begin{figure}
	\epsscale{1.0}
	\plotone{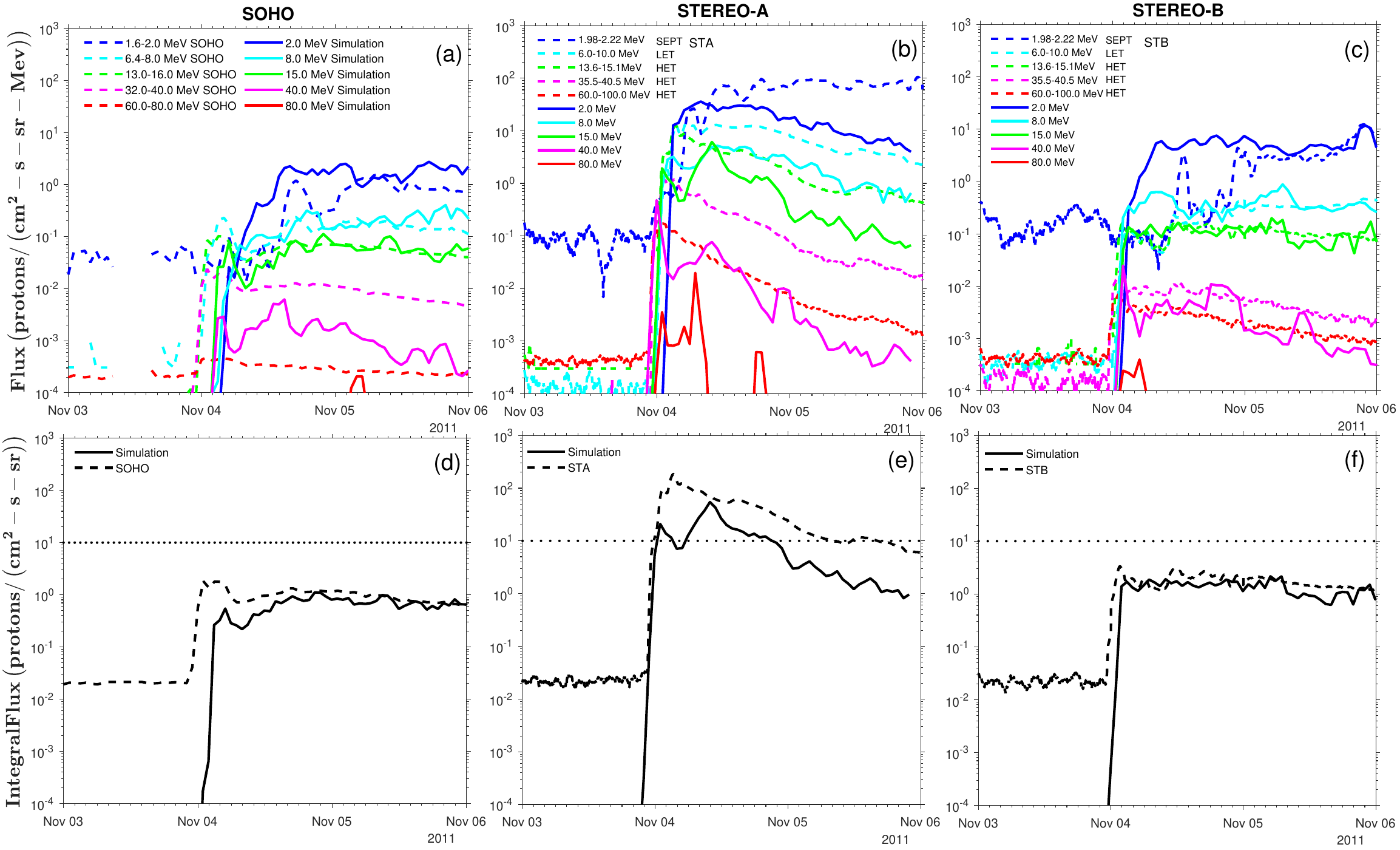}
	\caption{
		Time evolution of (Top) differential proton  flux between 2.0 MeV and 80.0 MeV at the loci of (a) Earth, (b) \textit{STEREO-A} and (c) \textit{STEREO-B},  and (Bottom)  integral flux for $>$10.0 MeV protons at the location of (d) Earth, (e) \textit{STEREO-A} and (f) \textit{STEREO-B} from the simulation (solid curves) and  observations (dash curves) for 2011 November 3 SEP event. The dotted line denotes the threshold flux  (10 proton flux units ((cm$^{2}$ s sr)$^{-1}$) for energies $>$ 10 MeV)  of a SEP event in the traditional NOAA definition. Data of observations at Earth are from ERNE instrument of \textit{SOHO} near the Sun-Earth L1 point.  Data of observations at \textit{STEREO-A} and  \textit{STEREO-B}  are from the SEPT, LET and HET instruments of \textit{STEREO-A} and \textit{STEREO-B}. 
		\label{fig:2011nov3_flux}}
\end{figure}

\subsection{2014 February 25 SEP Event}
On 2014 February 25 (day of year 56), \textit{STEREO-A} was located 152$^{\circ}$ in longitude west of Earth and at a heliocentric radial distance  of 0.99 au
while STEREO-B was east of Earth by
$\sim160^{\circ}$ and  at a heliocentric radial distance  of 1.06 au as seen in Figure \ref{fig:2014feb25location}(a). The   solar flare associated with the origin of the SEP event was located at S15E77, started around 00:41 UT and peaked at 00:49 UT with an estimated X4.9 class intensity as reported on the DONKI website.  A halo CME was reported in the  LASCO CME Catalog with a speed of 2147 $\rm km\;s^{-1}$. There was another prior CME with a speed of 790 $\rm km\;s^{-1}$ at 23:24 UT
on 2014 February 24. This slower CME was not associated with the  SEP event on 2014 February 25 \citep{2016ApJ...819...72L}. All of the spacecraft \textit{STEREO-A}, \textit{SOHO} and \textit{STEREO-B} observed this SEP event indicating a large  longitudinal extent.

Figure \ref{fig:2014feb25location}(b) illustrates the projection on the solar surface of \textit{STEREO-A} (STA), \textit{Earth} and \textit{STEREO-B} (STB). The footpoints of the field lines connecting to STEREO-A and STEREO-B are very close among them and separated in longitude by  $\sim 36^{\circ}$ from the solar flare site. The longitudinal distance between the footpoints of the field lines connecting to \textit{Earth} and the solar flare is larger $\sim$ 130$^{\circ}$.

The time evolution of the ellipsoid CME shock surfaces (orange surfaces) with magnetic field lines connecting to \textit{Earth} (green), \textit{STEREO-A} (red), and \textit{STEREO-B} (blue) is illustrated in Figure \ref{fig:2014feb25CME}. Throughout the propagation of the CME shock, \textit{STEREO-B} maintained a magnetic connection to the shock surface for the longest duration from 00:55:21 UT on 2014 February 25 to 08:24:21 UT on 2014 February 26 . Initially, only \textit{STEREO-B} was magnetically connected to the CME shock at 00:55:21 UT on 2014 February 25. Around 01:05:21 UT, \textit{STEREO-A} began to establish  magnetic connection with the shock surface and maintained the connection until 03:54:21 UT. \textit{Earth} subsequently established magnetic connection with the reconstructed CME shock at 01:24:35UT, but such connection was lost  around half an hour later at 01:54:21 UT. After 08:24:21 UT on 2014 February 26, none of the spacecraft were magnetically connected to the CME shock.

\begin{figure}
	\epsscale{0.9}
	\plotone{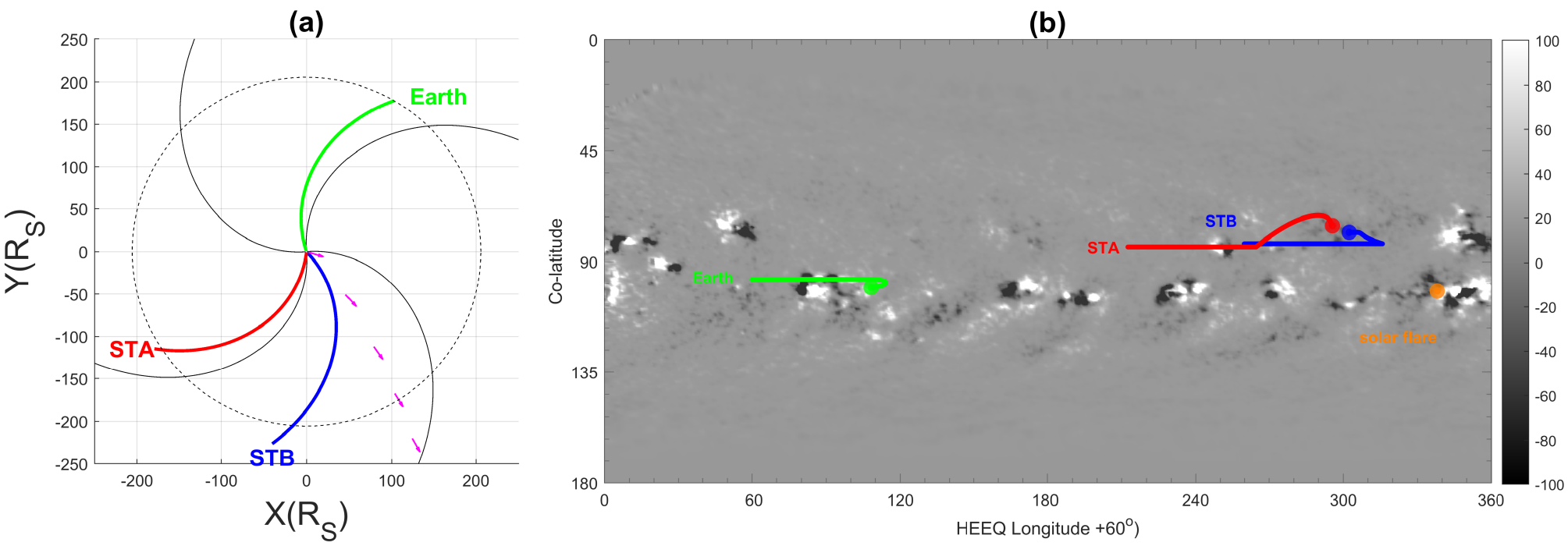}
	\caption{(a) Locations of \textit{STEREO-A}(STA), \textit{Earth} and \textit{STEREO-B} (STB) on 2014 February 25 in the equatorial plane as seen from the north, where $R_S$ is the solar radius. The magenta arrow denotes the moving direction of the CME shock; The blue, green, and red curves are the Parker spiral and coronal magnetic field lines that connect to the spacecraft.   The panel (b) shows the projection on the solar surface of \textit{STEREO-A}(STA), \textit{Earth} and \textit{STEREO-B} (STB). The gray background image is the strength of the measured photospheric magnetic field.  The red, green and blue dots mark the footpoints of the field lines connecting to \textit{STEREO-A}, \textit{Earth} and \textit{STEREO-B}, respectively. The orange dot is the site of solar flare. A shift ($60^{\circ}$) of the heliocentric Earth equatorial (HEEQ) coordinate system in longitude is used. Earth is at $60^{\circ}$ in longitude.
		\label{fig:2014feb25location}}
\end{figure}

\begin{figure}
	\epsscale{1.1}
	\plotone{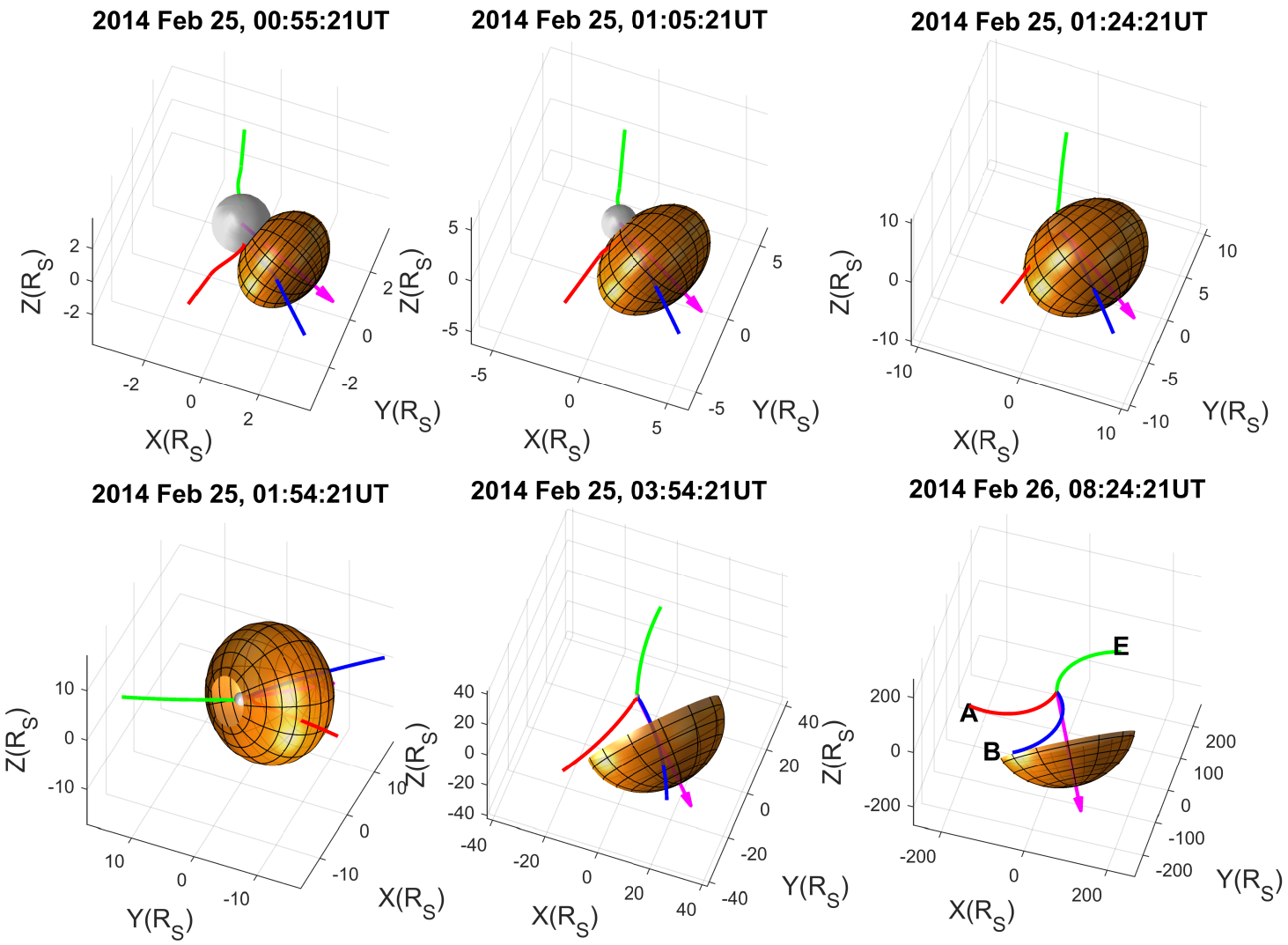}
	\caption{Time evolution of ellipsoid CME shock surfaces (orange color) on 2014 February 25 with magnetic field lines connecting to \textit{Earth} (green), \textit{STEREO-A} (STA) (red), and \textit{STEREO-B} (STB) (blue), where the Sun (gray sphere) is at the center. The purple arrow denotes the moving direction of the CME shock.
		\label{fig:2014feb25CME}}
\end{figure}

The rise in proton flux is observed by ACE, GOES, SOHO, STEREO-A and STEREO-B as shown in Figure \ref{fig:2014feb25flux}. For low-energy protons, the modeled fluxes agree with the observed ones at Earth and STEREO-B. The modeled proton fluxes at high energies  are overestimated  due to an overestimation of shock cutoff energy. The simulated proton intensities at STEREO-A display a delayed onset and a lower intensity with respect to observations as seen in Figures \ref{fig:2014feb25flux}(b) and \ref{fig:2014feb25flux}(e). Delays of the modeled onsets are also seen at Earth and STEREO-B in Figures \ref{fig:2014feb25flux}(d) and  \ref{fig:2014feb25flux}(f). We might be able to reduce such a delay by adjusting both parallel and perpendicular mean free paths, but we are not conducting a parameter search at this time due to computational constraints. At STEREO-A (Figure \ref{fig:2014feb25flux}(b)), the modeled differential flux for protons at low energies are lower than the observations because of the influence of shock-generated turbulence while the flux for protons at higher energies display an opposite behavior attributed to 
an overestimation of shock cutoff energy as discussed in previous events.

\begin{figure}
	\epsscale{1.2}
	\plotone{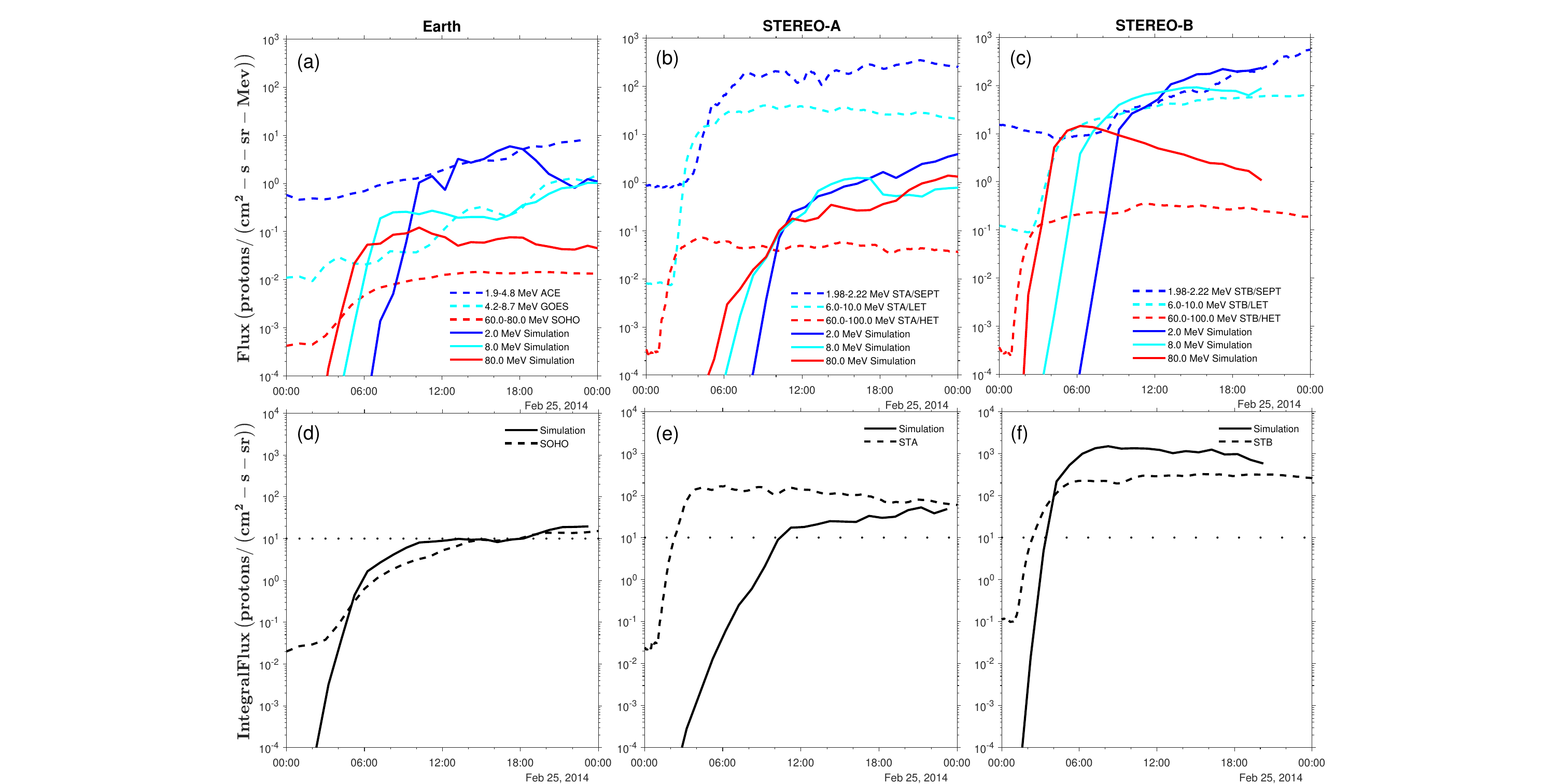}
	\caption{Time evolution of (Top) differential proton  flux on 2014 February 25 between 2.0 MeV and 80.0 MeV  at the loci of (a) Earth, (b) \textit{STEREO-A} and (c) \textit{STEREO-B},  and (Bottom)  integral flux for $>$10.0 MeV protons at the location of (d) Earth, (e) \textit{STEREO-A} and (f) textit{STEREO-B} from the simulation (solid curves) and  observations (dash curves). The dotted line denotes the threshold flux  (10 proton flux units ((cm$^{2}$ s sr)$^{-1}$) for energies $>$ 10 MeV)  of a SEP event in the traditional NOAA definition. Data of observations at Earth are from ACE, GOES and ERNE instrument of \textit{SOHO}.  Data of observations at \textit{STEREO-A} and  \textit{STEREO-B}  are from the SEPT, LET and HET instruments of \textit{STEREO-A} and \textit{STEREO-B}. 
		\label{fig:2014feb25flux}}
\end{figure}

\section{Summary and Discussion}\label{sec:disc}

In this study, we investigate the performance of our model by employing time-backward stochastic differential equations to derive the intensities of energetic particles during SEP events associated with fast CMEs (speeds exceeding 700 $\rm km \;s^{-1}$). The transport equation is utilized to describe the particle dynamics, incorporating a SEP source term consistent with the theory of diffusive shock acceleration by a CME shock front. The input for our model includes the CME shock reconstructed from coronagraph observations and the magnetic field configuration obtained from the PFSS model. The analytic model proposed by \citet{2013SoPh..285..391C,2017SpWea..15..464C} is adopted for CME propagation beyond the coronagraph field of view. It is important to note that the simulation described above is specific to SEP protons and does not include SEP electrons.

We applied our simulation code to analyze the SEP events that occurred on 2013 April 11, 2011 March 7, 2011 November 3 and 2014 February 25  to obtain the time profiles of proton flux and compare them with multi spacecraft observations. Overall, our simulation results demonstrate good agreement with the observations, particularly for protons with energies below 40.0 MeV. However, for protons above 40.0 MeV, near the cut-off frequency, our simulation tends to overestimate the observed proton flux in the case of the 2011 March 7 and 2014 February 25 SEP events. This discrepancy may be attributed to the fact that the particles in the low-energy channels have energies below the shock acceleration cutoff in the corona. On the other hand, the high-energy channels contain particles with energies above the shock cutoff.  It is important to note that when calculating the shock cutoff using Bohm diffusion, there is a possibility of overestimating the cutoff energy. At STEREO-A for SEP events on 2013 April 11 and 2011 March 7, the modeled proton intensities display a very different behavior compared to observations because of the large transport in longitude caused by the weak magnetic field.
For the 2011 November 3 SEP event, the intensity at Earth and STEREO-B align well with observations for low-energy channels, but they are underestimated for high-energy channels above 40 MeV \citep{Zhang_2023,Zhang:2023aL}. Nevertheless, simulations conducted at STEREO-A reproduce a lower proton intensity compared to the observed results.  \citet{Zhang_2023} pointed out that this phenomenon is attributed to the influence of shock-generated turbulence, which alters the diffusion coefficient along the magnetic field lines connected to STEREO-A. Simulation with different magnetic field models which can describe the magnetic field more realistically like MHD model and a horizontal current–current
sheet–source surface (HCCSSS) model developed at Stanford University \citep{1994SoPh..151...91Z,1995JGR...100...19Z} might improve the performance.

M.Z. acknowledges support from NASA Grants 80NSSC21K0004, 80NSSC20K0298, 80NSSC20K0286, 80NSSC19K1254. 

This CME catalog is generated and maintained at the CDAW Data Center by NASA and The Catholic University of America in cooperation with the Naval Research Laboratory. SOHO is a project of international cooperation between ESA and NASA. The data from the observation are downloaded from \url{http://spdf.gsfc.nasa.gov}. We acknowledge use of NASA/GSFC's Space Physics Data Facility's OMNIWeb (or CDAWeb or ftp) service, and OMNI data. The computation facility used to run the simulations in the paper was supported by the National Science Foundation under Grant No. CNS 2016818.

%
%
%


\begin{thebibliography}{}
\expandafter\ifx\csname natexlab\endcsname\relax\def\natexlab#1{#1}\fi
\providecommand{\url}[1]{\href{#1}{#1}}
\providecommand{\dodoi}[1]{doi:~\href{http://doi.org/#1}{\nolinkurl{#1}}}
\providecommand{\doeprint}[1]{\href{http://ascl.net/#1}{\nolinkurl{http://ascl.net/#1}}}
\providecommand{\doarXiv}[1]{\href{https://arxiv.org/abs/#1}{\nolinkurl{https://arxiv.org/abs/#1}}}

\bibitem[{{Baring}(1997)}]{baring1997diffusive}
{Baring}, M.~G. 1997, in Very High Energy Phenomena in the Universe; Moriond
  Workshop, ed. Y.~{Giraud-Heraud} \& J.~{Tran Thanh van}, 97.
\newblock \doarXiv{astro-ph/9711177}

\bibitem[{{Bieber} {et~al.}(1994){Bieber}, {Matthaeus}, {Smith}, {Wanner},
  {Kallenrode}, \& {Wibberenz}}]{Bieberetal1994}
{Bieber}, J.~W., {Matthaeus}, W.~H., {Smith}, C.~W., {et~al.} 1994, \apj, 420,
  294, \dodoi{10.1086/173559}

\bibitem[{Cheng {et~al.}(2023)Cheng, Zhang, Lario, Balmaceda, Kwon, \&
  Cohen}]{Cheng_2023}
Cheng, L., Zhang, M., Lario, D., {et~al.} 2023, The Astrophysical Journal, 943,
  134, \dodoi{10.3847/1538-4357/acac21}

\bibitem[{{Corona-Romero} {et~al.}(2013){Corona-Romero}, {Gonzalez-Esparza}, \&
  {Aguilar-Rodriguez}}]{2013SoPh..285..391C}
{Corona-Romero}, P., {Gonzalez-Esparza}, J.~A., \& {Aguilar-Rodriguez}, E.
  2013, \solphys, 285, 391, \dodoi{10.1007/s11207-012-0103-9}

\bibitem[{{Corona-Romero} {et~al.}(2017){Corona-Romero}, {Gonzalez-Esparza},
  {Perez-Alanis}, {Aguilar-Rodriguez}, {de-la-Luz}, \&
  {Mejia-Ambriz}}]{2017SpWea..15..464C}
{Corona-Romero}, P., {Gonzalez-Esparza}, J.~A., {Perez-Alanis}, C.~A., {et~al.}
  2017, Space Weather, 15, 464, \dodoi{10.1002/2016SW001489}

\bibitem[{{Dr\"{o}ge}(1994)}]{Droge1994}
{Dr\"{o}ge}, W. 1994, \apjs, 90, 567, \dodoi{10.1086/191876}

\bibitem[{{Dr{\"o}ge} {et~al.}(2010){Dr{\"o}ge}, {Kartavykh}, {Klecker}, \&
  {Kovaltsov}}]{Drogeetal2010}
{Dr{\"o}ge}, W., {Kartavykh}, Y.~Y., {Klecker}, B., \& {Kovaltsov}, G.~A. 2010,
  \apj, 709, 912, \dodoi{10.1088/0004-637X/709/2/912}

\bibitem[{{Drury}(1983)}]{Drury1983}
{Drury}, L.~O. 1983, Reports on Progress in Physics, 46, 973,
  \dodoi{10.1088/0034-4885/46/8/002}

\bibitem[{Fitzpatrick(2014)}]{fitzpatrick2014plasma}
Fitzpatrick, R. 2014, Plasma physics: an introduction (Crc Press), 220

\bibitem[{{Giacalone} {et~al.}(2000){Giacalone}, {Jokipii}, \&
  {Mazur}}]{giacalone2000small}
{Giacalone}, J., {Jokipii}, J.~R., \& {Mazur}, J.~E. 2000, \apjl, 532, L75,
  \dodoi{10.1086/312564}

\bibitem[{{Gopalswamy} {et~al.}(2009){Gopalswamy}, {Yashiro}, {Michalek},
  {Stenborg}, {Vourlidas}, {Freeland}, \& {Howard}}]{2009EM&P..104..295G}
{Gopalswamy}, N., {Yashiro}, S., {Michalek}, G., {et~al.} 2009, Earth Moon and
  Planets, 104, 295, \dodoi{10.1007/s11038-008-9282-7}

\bibitem[{Gómez-Herrero {et~al.}(2015)Gómez-Herrero, Dresing, Klassen, Heber,
  Lario, Agueda, Malandraki, Blanco, Rodríguez-Pacheco, \&
  Banjac}]{Gomez_2015}
Gómez-Herrero, R., Dresing, N., Klassen, A., {et~al.} 2015, The Astrophysical
  Journal, 799, 55, \dodoi{10.1088/0004-637X/799/1/55}

\bibitem[{{Heras} {et~al.}(1995){Heras}, {Sanahuja}, {Lario}, {Smith},
  {Detman}, \& {Dryer}}]{heras1995three}
{Heras}, A.~M., {Sanahuja}, B., {Lario}, D., {et~al.} 1995, \apj, 445, 497,
  \dodoi{10.1086/175714}

\bibitem[{{Heras} {et~al.}(1992){Heras}, {Sanahuja}, {Smith}, {Detman}, \&
  {Dryer}}]{heras1992influence}
{Heras}, A.~M., {Sanahuja}, B., {Smith}, Z.~K., {Detman}, T., \& {Dryer}, M.
  1992, \apj, 391, 359, \dodoi{10.1086/171351}

\bibitem[{{Hu} {et~al.}(2017){Hu}, {Li}, {Ao}, {Zank}, \&
  {Verkhoglyadova}}]{Huetal2017}
{Hu}, J., {Li}, G., {Ao}, X., {Zank}, G.~P., \& {Verkhoglyadova}, O. 2017,
  Journal of Geophysical Research (Space Physics), 122, 10,938,
  \dodoi{10.1002/2017JA024077}

\bibitem[{{Kabin}(2001)}]{Kabin2001JPlPh..66..259K}
{Kabin}, K. 2001, Journal of Plasma Physics, 66, 259,
  \dodoi{10.1017/S0022377801001295}

\bibitem[{{Kallenrode}(1993)}]{Kallenrode1993}
{Kallenrode}, M.-B. 1993, \jgr, 98, 19037, \dodoi{10.1029/93JA02079}

\bibitem[{{Kallenrode} \& {Wibberenz}(1997)}]{KallenrodeWibberenz1997}
{Kallenrode}, M.-B., \& {Wibberenz}, G. 1997, \jgr, 102, 22311,
  \dodoi{10.1029/97JA01677}

\bibitem[{{Kozarev} {et~al.}(2013){Kozarev}, {Evans}, {Schwadron}, {Dayeh},
  {Opher}, {Korreck}, \& {van der Holst}}]{Kozarevetal2013}
{Kozarev}, K.~A., {Evans}, R.~M., {Schwadron}, N.~A., {et~al.} 2013, \apj, 778,
  43, \dodoi{10.1088/0004-637X/778/1/43}

\bibitem[{Kwon {et~al.}(2014)Kwon, Zhang, \& Olmedo}]{kwon2014new}
Kwon, R.-Y., Zhang, J., \& Olmedo, O. 2014, The Astrophysical Journal, 794,
  148, \dodoi{10.1088/0004-637x/794/2/148}

\bibitem[{{Lario} {et~al.}(2014){Lario}, {Raouafi}, {Kwon}, {Zhang},
  {G{\'o}mez-Herrero}, {Dresing}, \& {Riley}}]{2014ApJ...797....8L}
{Lario}, D., {Raouafi}, N.~E., {Kwon}, R.~Y., {et~al.} 2014, \apj, 797, 8,
  \dodoi{10.1088/0004-637X/797/1/8}

\bibitem[{{Lario} {et~al.}(1998){Lario}, {Sanahuja}, \&
  {Heras}}]{lario1998energetic}
{Lario}, D., {Sanahuja}, B., \& {Heras}, A.~M. 1998, \apj, 509, 415,
  \dodoi{10.1086/306461}

\bibitem[{{Lario} {et~al.}(2016){Lario}, {Kwon}, {Vourlidas}, {Raouafi},
  {Haggerty}, {Ho}, {Anderson}, {Papaioannou}, {G{\'o}mez-Herrero}, {Dresing},
  \& {Riley}}]{2016ApJ...819...72L}
{Lario}, D., {Kwon}, R.~Y., {Vourlidas}, A., {et~al.} 2016, \apj, 819, 72,
  \dodoi{10.3847/0004-637X/819/1/72}

\bibitem[{le~Roux \& Webb(2009)}]{le_Roux_2009}
le~Roux, J.~A., \& Webb, G.~M. 2009, The Astrophysical Journal, 693, 534,
  \dodoi{10.1088/0004-637x/693/1/534}

\bibitem[{Leblanc {et~al.}(1998)Leblanc, Dulk, \&
  Bougeret}]{leblanc1998tracing}
Leblanc, Y., Dulk, G.~A., \& Bougeret, J.-L. 1998, Solar Physics, 183, 165,
  \dodoi{10.1023/A:1005049730506}

\bibitem[{{Lee}(2005)}]{Lee2005}
{Lee}, M.~A. 2005, \apjs, 158, 38, \dodoi{10.1086/428753}

\bibitem[{Lee(2005)}]{Lee_2005}
Lee, M.~A. 2005, The Astrophysical Journal Supplement Series, 158, 38,
  \dodoi{10.1086/428753}

\bibitem[{{Li} {et~al.}(2003){Li}, {Zank}, \& {Rice}}]{Lietal2003}
{Li}, G., {Zank}, G.~P., \& {Rice}, W.~K.~M. 2003, Journal of Geophysical
  Research (Space Physics), 108, 1082, \dodoi{10.1029/2002JA009666}

\bibitem[{{Luhmann} {et~al.}(2010){Luhmann}, {Ledvina}, {Odstrcil}, {Owens},
  {Zhao}, {Liu}, \& {Riley}}]{Luhmannetal2010}
{Luhmann}, J.~G., {Ledvina}, S.~A., {Odstrcil}, D., {et~al.} 2010, Advances in
  Space Research, 46, 1, \dodoi{10.1016/j.asr.2010.03.011}

\bibitem[{{Marsh} {et~al.}(2015){Marsh}, {Dalla}, {Dierckxsens}, {Laitinen}, \&
  {Crosby}}]{Marshetal2015}
{Marsh}, M.~S., {Dalla}, S., {Dierckxsens}, M., {Laitinen}, T., \& {Crosby},
  N.~B. 2015, Space Weather, 13, 386, \dodoi{10.1002/2014SW001120}

\bibitem[{{Mewaldt} {et~al.}(2008){Mewaldt}, {Cohen}, {Cook}, {Cummings},
  {Davis}, {Geier}, {Kecman}, {Klemic}, {Labrador}, {Leske}, {Miyasaka},
  {Nguyen}, {Ogliore}, {Stone}, {Radocinski}, {Wiedenbeck}, {Hawk}, {Shuman},
  {von Rosenvinge}, \& {Wortman}}]{mewaldt2008low}
{Mewaldt}, R.~A., {Cohen}, C.~M.~S., {Cook}, W.~R., {et~al.} 2008, \ssr, 136,
  285, \dodoi{10.1007/s11214-007-9288-x}

\bibitem[{{M{\"u}ller-Mellin} {et~al.}(2008){M{\"u}ller-Mellin},
  {B{\"o}ttcher}, {Falenski}, {Rode}, {Duvet}, {Sanderson}, {Butler},
  {Johlander}, \& {Smit}}]{2008SSRv..136..363M}
{M{\"u}ller-Mellin}, R., {B{\"o}ttcher}, S., {Falenski}, J., {et~al.} 2008,
  \ssr, 136, 363, \dodoi{10.1007/s11214-007-9204-4}

\bibitem[{{Ng} \& {Reames}(1994)}]{NgReames1994}
{Ng}, C.~K., \& {Reames}, D.~V. 1994, \apj, 424, 1032, \dodoi{10.1086/173954}

\bibitem[{{Ng} {et~al.}(2003){Ng}, {Reames}, \& {Tylka}}]{Ngetal2003}
{Ng}, C.~K., {Reames}, D.~V., \& {Tylka}, A.~J. 2003, \apj, 591, 461,
  \dodoi{10.1086/375293}

\bibitem[{{Nitta} {et~al.}(2013){Nitta}, {Aschwanden}, {Boerner}, {Freeland},
  {Lemen}, \& {Wuelser}}]{2013SoPh..288..241N}
{Nitta}, N.~V., {Aschwanden}, M.~J., {Boerner}, P.~F., {et~al.} 2013, \solphys,
  288, 241, \dodoi{10.1007/s11207-013-0307-7}

\bibitem[{Park {et~al.}(2013)Park, Innes, Bucik, \& Moon}]{Park_2013}
Park, J., Innes, D.~E., Bucik, R., \& Moon, Y.-J. 2013, The Astrophysical
  Journal, 779, 184, \dodoi{10.1088/0004-637X/779/2/184}

\bibitem[{{Prise} {et~al.}(2014){Prise}, {Harra}, {Matthews}, {Long}, \&
  {Aylward}}]{2014SoPh..289.1731P}
{Prise}, A.~J., {Harra}, L.~K., {Matthews}, S.~A., {Long}, D.~M., \& {Aylward},
  A.~D. 2014, \solphys, 289, 1731, \dodoi{10.1007/s11207-013-0435-0}

\bibitem[{Qin {et~al.}(2006)Qin, Zhang, \& Dwyer}]{qin2006effect}
Qin, G., Zhang, M., \& Dwyer, J.~R. 2006, Journal of Geophysical Research:
  Space Physics, 111, \dodoi{https://doi.org/10.1029/2005JA011512}

\bibitem[{{Rice} {et~al.}(2003){Rice}, {Zank}, \& {Li}}]{Riceetal2003}
{Rice}, W.~K.~M., {Zank}, G.~P., \& {Li}, G. 2003, Journal of Geophysical
  Research (Space Physics), 108, 1369, \dodoi{10.1029/2002JA009756}

\bibitem[{{Ruffolo}(1995)}]{Ruffolo1995}
{Ruffolo}, D. 1995, \apj, 442, 861, \dodoi{10.1086/175489}

\bibitem[{{Thompson}(1962)}]{Thompson1962}
{Thompson}, W.~B. 1962, {An Introduction to Plasma Physics} (Addison Wesley),
  86--95

\bibitem[{{Torsti} {et~al.}(1995){Torsti}, {Valtonen}, {Lumme}, {Peltonen},
  {Eronen}, {Louhola}, {Riihonen}, {Schultz}, {Teittinen}, {Ahola}, {Holmlund},
  {Kelh{\"a}}, {Lepp{\"a}l{\"a}}, {Ruuska}, \&
  {Str{\"o}mmer}}]{1995SoPh..162..505T}
{Torsti}, J., {Valtonen}, E., {Lumme}, M., {et~al.} 1995, \solphys, 162, 505,
  \dodoi{10.1007/BF00733438}

\bibitem[{{von Rosenvinge} {et~al.}(2008){von Rosenvinge}, {Reames}, {Baker},
  {Hawk}, {Nolan}, {Ryan}, {Shuman}, {Wortman}, {Mewaldt}, {Cummings}, {Cook},
  {Labrador}, {Leske}, \& {Wiedenbeck}}]{2008SSRv..136..391V}
{von Rosenvinge}, T.~T., {Reames}, D.~V., {Baker}, R., {et~al.} 2008, \ssr,
  136, 391, \dodoi{10.1007/s11214-007-9300-5}

\bibitem[{{Wijsen} {et~al.}(2022){Wijsen}, {Aran}, {Scolini}, {Lario},
  {Afanasiev}, {Vainio}, {Sanahuja}, {Pomoell}, \& {Poedts}}]{Wijsen2022}
{Wijsen}, N., {Aran}, A., {Scolini}, C., {et~al.} 2022, A\&A, 659, A187,
  \dodoi{10.1051/0004-6361/202142698}

\bibitem[{{Yashiro} {et~al.}(2004){Yashiro}, {Gopalswamy}, {Michalek}, {St.
  Cyr}, {Plunkett}, {Rich}, \& {Howard}}]{2004JGRA..109.7105Y}
{Yashiro}, S., {Gopalswamy}, N., {Michalek}, G., {et~al.} 2004, Journal of
  Geophysical Research (Space Physics), 109, A07105,
  \dodoi{10.1029/2003JA010282}

\bibitem[{Zank {et~al.}(2000)Zank, Rice, \& Wu}]{zank2000particle}
Zank, G.~P., Rice, W. K.~M., \& Wu, C.~C. 2000, Journal of Geophysical
  Research: Space Physics, 105, 25079,
  \dodoi{https://doi.org/10.1029/1999JA000455}

\bibitem[{Zhang \& Cheng(2023)}]{Zhang:2023aL}
Zhang, M., \& Cheng, L. 2023, PoS, ICRC2023, 1275, \dodoi{10.22323/1.444.1275}

\bibitem[{Zhang {et~al.}(2023)Zhang, Cheng, Zhang, Riley, Kwon, Lario,
  Balmaceda, \& Pogorelov}]{Zhang_2023}
Zhang, M., Cheng, L., Zhang, J., {et~al.} 2023, The Astrophysical Journal
  Supplement Series, 266, 35, \dodoi{10.3847/1538-4365/accb8e}

\bibitem[{Zhang {et~al.}(2009)Zhang, Qin, \& Rassoul}]{zhang2009propagation}
Zhang, M., Qin, G., \& Rassoul, H. 2009, The Astrophysical Journal, 692, 109,
  \dodoi{10.1088/0004-637x/692/1/109}

\bibitem[{Zhang \& Zhao(2017)}]{zhang2017precipitation}
Zhang, M., \& Zhao, L. 2017, The Astrophysical Journal, 846, 107,
  \dodoi{10.3847/1538-4357/aa86a8}

\bibitem[{Zhao \& Zhang(2018)}]{Zhao_2018}
Zhao, L., \& Zhang, M. 2018, The Astrophysical Journal, 859, L29,
  \dodoi{10.3847/2041-8213/aac6cf}

\bibitem[{{Zhao} \& {Hoeksema}(1994)}]{1994SoPh..151...91Z}
{Zhao}, X., \& {Hoeksema}, J.~T. 1994, \solphys, 151, 91,
  \dodoi{10.1007/BF00654084}

\bibitem[{{Zhao} \& {Hoeksema}(1995)}]{1995JGR...100...19Z}
---. 1995, \jgr, 100, 19, \dodoi{10.1029/94JA02266}

\end{thebibliography}

\listofchanges
\end{document}